\documentclass[aps,nofootinbib,pre,reprint,twocolumn,notitlepage,superscriptaddress]{revtex4}
\usepackage{graphicx}
\usepackage{epstopdf}
\usepackage{bm}
\usepackage{amssymb}
\usepackage{psfrag}
\usepackage{color}
\usepackage{amsbsy}      
\usepackage{amsmath,amsthm}
\usepackage{mathtools}
\usepackage{dsfont}
\usepackage{makecell}
\usepackage{slashbox}
\usepackage{tcolorbox}
\usepackage{soul,color}
\usepackage{url}
\usepackage{etoolbox}
\patchcmd{\section}
  {\centering}
  {\raggedright}
  {}
  {}
\patchcmd{\subsection}
  {\centering}
  {\raggedright}
  {}
  {}
\graphicspath{{./figs/}}
\usepackage{newfloat}
\DeclareFloatingEnvironment[name={Fig. D}]{suppfigure}

\usepackage[sort&compress]{natbib} 

\def\dd{\mbox{d}}

\newcommand*\widebar[1]{%
   \hbox{%
     \vbox{%
       \hrule height 0.5pt 
       \kern0.5ex
       \hbox{%
         \kern-0.1em
         \ensuremath{#1}%
         \kern-0.1em
       }%
     }%
   }%
}

\def\ie{\textit{i.e.}}
\def\eg{\textit{e.g.}}

\begin{document}

\title{Using excess deaths and testing statistics to improve estimates
  of COVID-19 mortalities}

\author{Lucas B\"{o}ttcher}
\email{lucasb@ucla.edu}
\affiliation{Dept.~of Computational Medicine, UCLA, Los Angeles, CA 90095-1766}
\author{Maria R. D'Orsogna}
\email{dorsogna@csun.edu}
\affiliation{Dept.~of Mathematics, California State University at Northridge, Los Angeles, CA 91330-8313}
\affiliation{Dept.~of Computational Medicine, UCLA, Los Angeles, CA 90095-1766}
\author{Tom Chou}
\email{tomchou@ucla.edu}
\affiliation{Dept.~of Computational Medicine, UCLA, Los Angeles, CA 90095-1766}
\affiliation{Dept.~of Mathematics, UCLA, Los Angeles, CA 90095-1555}
\date{\today}
%
\begin{abstract}
Factors such as non-uniform definitions of mortality, uncertainty in
disease prevalence, and biased sampling complicate the quantification
of fatality during an epidemic. Regardless of the employed fatality
measure, the infected population and the number of infection-caused
deaths need to be consistently estimated for comparing mortality
across regions. We combine historical and current mortality data, a
statistical testing model, and an SIR epidemic model, to improve
estimation of mortality. We find that the average excess death across
the entire US is 13$\%$ higher than the number of reported COVID-19
deaths. In some areas, such as New York City, the number of weekly
deaths is about eight times higher than in previous years. Other
countries such as Peru, Ecuador, Mexico, and Spain exhibit excess
deaths significantly higher than their reported COVID-19
deaths. Conversely, we find negligible or negative excess deaths for
part and all of 2020 for Denmark, Germany, and Norway.
\end{abstract}
\maketitle
\section*{Introduction}
The novel severe acute respiratory syndrome coronavirus 2 (SARS-CoV-2)
first identified in Wuhan, China in December 2019 quickly spread
across the globe, leading to the declaration of a pandemic on March
11, 2020~\cite{WHOpandemic}.  The emerging disease was termed
COVID-19.  As of this January 2020 writing, more than 86 million
people have been infected, and more than 1.8 million deaths from
COVID-19 in more than 218 countries~\cite{corona1} have been
confirmed. About 61 million people have recovered globally.

Properly estimating the severity of any infectious disease is crucial
for identifying near-future scenarios, and designing intervention
strategies. This is especially true for SARS-CoV-2 given the relative
ease with which it spreads, due to long incubation periods,
asymptomatic carriers, and stealth
transmissions~\cite{he2020temporal}.  Most measures of severity are
derived from the number of deaths, the number of confirmed and
unconfirmed infections, and the number of secondary cases generated by
a single primary infection, to name a few.  Measuring these
quantities, determining how they evolve in a population, and how they
are to be compared across groups, and over time, is challenging due to
many confounding variables and uncertainties.

For example, quantifying COVID-19 deaths across jurisdictions must
take into account the existence of different protocols in assigning
cause of death, cataloging co-morbidities \cite{CDC_comorbidity}, and
lag time reporting~\cite{BBC_deaths}. Inconsistencies also arise in
the way deaths are recorded, especially when COVID-19 is not the
direct cause of death, rather a co-factor leading to complications
such as pneumonia and other respiratory
ailments~\cite{beaney2020excess}. In Italy, the clinician's best
judgment is called upon to classify the cause of death of an untested
person who manifests COVID-19 symptoms.  In some cases, such persons
are given postmortem tests, and if results are positive, added to the
statistics. Criteria vary from region to
region~\cite{onder2020case}. In Germany, postmortem testing is not
routinely employed, possibly explaining the large difference in
mortality between the two countries. In the US, current guidelines
state that if typical symptoms are observed, the patient's death can
be registered as due to COVID-19 even without a positive
test~\cite{CDC_death_definition}.  Certain jurisdictions will list
dates on which deaths actually occurred, others list dates on which
they were reported, leading to potential lag-times. Other countries
tally COVID-19 related deaths only if they occur in hospital settings,
while others also include those that occur in private and/or nursing
homes.

In addition to the difficulty in obtaining accurate and uniform
fatality counts, estimating the prevalence of the disease is also a
challenging task. Large-scale testing of a population where a fraction
of individuals is infected, relies on unbiased sampling, reliable
tests, and accurate recording of results. One of the main sources of
systematic bias arises from the tested subpopulation: due to shortages
in testing resources, or in response to public health guidelines,
COVID-19 tests have more often been conducted on symptomatic persons,
the elderly, front-line workers and/or those returning from
hot-spots. Such non-random testing overestimates the infected fraction
of the population.

Different types of tests also probe different infected
subpopulations. Tests based on reverse-transcription polymerase chain
reaction (RT-PCR), whereby viral genetic material is detected
primarily in the upper respiratory tract and amplified, probe
individuals who are actively infected. Serological tests (such as
enzyme-linked immunosorbent assay, ELISA) detect antiviral antibodies
and thus measure individuals who have been infected, including those
who have recovered.

Finally, different types of tests exhibit significantly different
``Type I'' (false positive) and ``Type II'' (false negative) error
rates. The accuracy of RT-PCR tests depends on viral load which may be
too low to be detected in individuals at the early stages of the
infection, and may also depend on which sampling site in the body is
chosen.  Within serological testing, the kinetics of antibody response
are still largely unknown and it is not possible to determine if and
for how long a person may be immune from reinfection. Instrumentation
errors and sample contamination may also result in a considerable
number of false positives and/or false negatives. These errors
confound the inference of the infected fraction. Specifically, at low
prevalence, Type I false positive errors can significantly bias the 
estimation of the IFR.

Other quantities that are useful in tracking the dynamics of a
pandemic include the number of recovered individuals, tested, or
untested. These quantities may not be easily inferred from data and
need to be estimated from fitting mathematical models such as SIR-type
ODEs~\cite{keeling2011modeling}, age-structured
PDEs~\cite{bottcher2020case}, or network/contact
models~\cite{bottcher2017critical,bottcher2020unifying,pastor2015epidemic}.

Administration of tests and estimation of all quantities above can
vary widely across jurisdictions, making it difficult to properly
compare numbers across them. In this paper, we incorporate excess
death data, testing statistics, and mathematical modeling to
self-consistently compute and compare mortality across different
jurisdictions. In particular, we will use excess mortality
statistics~\cite{faust2020comparison,woolf2020excess,kontis2020magnitude}
to infer the number of COVID-19-induced deaths across different
regions. We then present a statistical testing model to estimate
jurisdiction-specific infected fractions and mortalities, their
uncertainty, and their dependence on testing bias and errors.  Our
statistical analyses and source codes are available at~\cite{GitHub}.

\section*{Methods}
\subsection*{Mortality measures}

Many different fatality rate measures have been defined to quantify
epidemic outbreaks~\cite{who_cfr_ifr}. One of the most common is the
case fatality ratio (${\rm CFR}$) defined as the ratio between the
number of confirmed ``infection-caused'' deaths $D_{\rm c}$ in a
specified time window and the number of infections $N_{\rm c}$
confirmed within the same time window, ${\rm CFR} = D_{\rm c}/N_{\rm
  c}$~\cite{xu2020pathological}. Depending on how deaths $D_{\rm c}$
are counted and how infected individuals $N_{\rm c}$ are defined, the
operational CFR may vary.  It may even exceed one, unless all deaths
are tested and included in $N_{\rm c}$.

Another frequently used measure is the infection fatality ratio (IFR)
defined as the true number of ``infection-caused'' deaths $D = D_{\rm
  c} + D_{\rm u}$ divided by the actual number of cumulative
infections to date, $N_{\rm c} + N_{\rm u}$. Here, $D_{\rm u}$ is the
number of unreported infection-caused deaths within a specified
period, and $N_{\rm u}$ denotes the untested or unreported infections
during the same period.  Thus, ${\rm IFR}= D/(N_{\rm c}+N_{\rm u})$.

One major issue of both CFR and IFR is that they do not account for
the time delay between infection and resolution. Both measures may be
quite inaccurate early in an outbreak when the number of cases grows
faster than the number of deaths and
recoveries~\cite{bottcher2020case}.  An alternative measure that
avoids case-resolution delays is the confirmed resolved mortality
$M=D_{\rm c}/(D_{\rm c}+R_{\rm c})$~\cite{bottcher2020case}, where
$R_{\rm c}$ is the cumulative number of confirmed recovered cases
evaluated in the same specified time window over which $D_{\rm c}$ is
counted.  One may also define the true resolved mortality via
$\mathcal{M}= D/(D + R)$, the proportion of the actual number of
deaths relative to the total number of deaths and recovered
individuals during a specified time period. If we decompose $R =
R_{\rm c} + R_{\rm u}$, where $R_{\rm c}$ are the confirmed and
$R_{\rm u}$, the unreported recovered cases, $\mathcal{M}= (D_{\rm
  c}+D_{\rm u})/(D_{\rm c}+D_{\rm u} + R_{\rm c}+R_{\rm u})$.
The total confirmed population is defined as 
%
%
$N_{\rm c} = D_{\rm c} + R_{\rm c} + I_{\rm c}$, where $I_{\rm c}$ the
number of living confirmed infecteds.  Applying these definitions to
any specified time period (typically from the ``start'' of an epidemic
to the date with the most recent case numbers), we observe that
$\mathrm{CFR} \leq M$ and $\mathrm{IFR} \leq {\cal M}$.  After the
epidemic has long past, when the number of currently infected
individuals $I$ approach zero, the two fatality ratios and mortality
measures converge if the component quantities are defined and measured
consistently, $\lim_{t \to \infty} \mathrm{CFR}(t) = \lim_{t \to
  \infty} M(t) $ and $\lim_{t \to \infty}\mathrm{IFR}(t) = \lim_{t \to
  \infty} \mathcal{M}(t)$~\cite{bottcher2020case}.

The mathematical definitions of the four basic mortality measures $Z =
{\rm CFR, IFR}, M, \cal M$ defined above are given in
Table~\ref{tab:mortality_measures} and fall into two categories,
confirmed and total. Confirmed measures (CFR and $M$) rely only on
positive test counts, while total measures (IFR and ${\cal M}$) rely
on projections to estimate the number of infected persons in the total
population $N$.
\begin{table*}[htb]
\renewcommand*{\arraystretch}{2.8}
\begin{tabular}{|c|c|c||c|}\hline
\backslashbox{Subpopulation}{Measure $Z$}
& \makebox[4em]{Fatality Ratios} & \makebox[6em]{Resolved Mortality} 
& \makebox[18em]{Excess Death Indices
}
\\\hline\hline
Confirmed & \,\,\, $\displaystyle{ 
{\rm CFR}= \frac{D_{\rm c}}{N_{\rm c}}}$\,\, & 
$\displaystyle M=\frac{D_{\rm c}} {D_{\rm c}+R_{\rm c}}$  & \,\,$D_{\rm e}$  per 100,000:   
$\displaystyle{ \frac{D_{\rm c}+D_{\rm u}}{100,000}}$ \,\, \\[3pt]\hline
Total &  \,\,$\displaystyle {\rm IFR}=\frac{D_{\rm c} + D_{\rm u}}{ N_{\rm c} 
+ N_{\rm u}}$\,\, &  
\,\, $\displaystyle {\cal M}=\frac{D_{\rm c}+D_{\rm u}}
{D_{\rm c} + D_{\rm u} + R_{\rm c} + R_{\rm u}}$\,\, 
& relative: $\displaystyle r = 
\frac{\sum_{i}\left[d^{(0)}(i) - {\frac 1 J} \sum_{j}^{J} d^{(j)}(i)\right]}
{{\frac 1 J}\sum_{j}^{J} \sum_{i}d^{(j)}(i)}$
\\[3pt]\hline
\end{tabular}
\vspace{1mm}
\caption{\textbf{Definitions of mortality measures.} Quantities with
  subscript ``c'' and ``u'' denote confirmed (\ie, positively tested)
  and unconfirmed populations. For instance, $D_{\rm c}$, $R_{\rm c}$,
  and $N_{\rm c}$ denote the total number of confirmed dead,
  recovered, and infected individuals, respectively. $d^{(j)}(i)$ is
  the number of individuals who have died in the $i^{\rm th}$ time
  window (\textit{e.g.}, day, week) of the $j^{\rm th}$ previous
  year. The mean number of excess deaths between the periods $k_{\rm
    s}$ and $k$ this year $\bar{D}_{\rm e}$ is thus $\sum_{i=k_{\rm
      s}}^k \left[d^{(0)}(i)- {\frac 1 J}\sum_{j=1}^{J}
    d^{(j)}(i)\right]$.  Where the total number of infection-caused
  deaths $D_{\rm c}+D_{\rm u}$ appears,it can be estimated using the
  excess deaths $\bar{D}_{\rm e}$ over as detailed in the main text.
  We have also included raw death numbers/100,000 and the mean excess
  deaths $r$ relative to the mean number of deaths over the same
  period of time from past years (see Eqs.~\eqref{DK_STATS}).}
\label{tab:mortality_measures}
\end{table*}
Of the measures listed in Table~\ref{tab:mortality_measures}, the
fatality ratio CFR and confirmed resolved mortality $M$ do not require
estimates of unreported infections, recoveries, and deaths and can be
directly derived from the available confirmed counts $D_{\rm c}$,
$N_{\rm c}$, and $R_{\rm c}$~\cite{dong2020interactive}.  Estimation
of IFR and the true resolved mortality $\cal M$ requires the
additional knowledge on the unconfirmed quantities $D_{\rm u}, N_{\rm
  u}$, and $R_{\rm u}$. We describe the possible ways to estimate
these quantities, along with the associated sources of bias and
uncertainty below.

\subsection*{Excess deaths data}

An unbiased way to estimate $D = D_{\rm c} + D_{\rm u}$, the
cumulative number of deaths, is to compare total deaths within a time
window in the current year to those in the same time window of
previous years, before the pandemic. If the epidemic is widespread and
has appreciable fatality, one may reasonably expect that the excess
deaths can be attributed to the
pandemic~\cite{USData,SpainData,EnglandData,SwitzerlandData,Istat}.
Within each affected region, these ``excess'' deaths $D_{\rm e}$
relative to ``historical'' deaths, are independent of testing
limitations and do not suffer from highly variable definitions of
virus-induced death. Thus, within the context of the COVID-19
pandemic, $D_{\rm e}$ is a more inclusive measure of virus-induced
deaths than $D_{\rm c}$ and can be used to estimate the total number
of deaths, $D_{\rm e} \simeq D_{\rm_c} + D_{\rm u}$. Moreover, using
data from multiple past years, one can also estimate the uncertainty
in $D_{\rm e}$.
\begin{figure*}[htb]
\centering
\includegraphics[width = 0.98\textwidth]{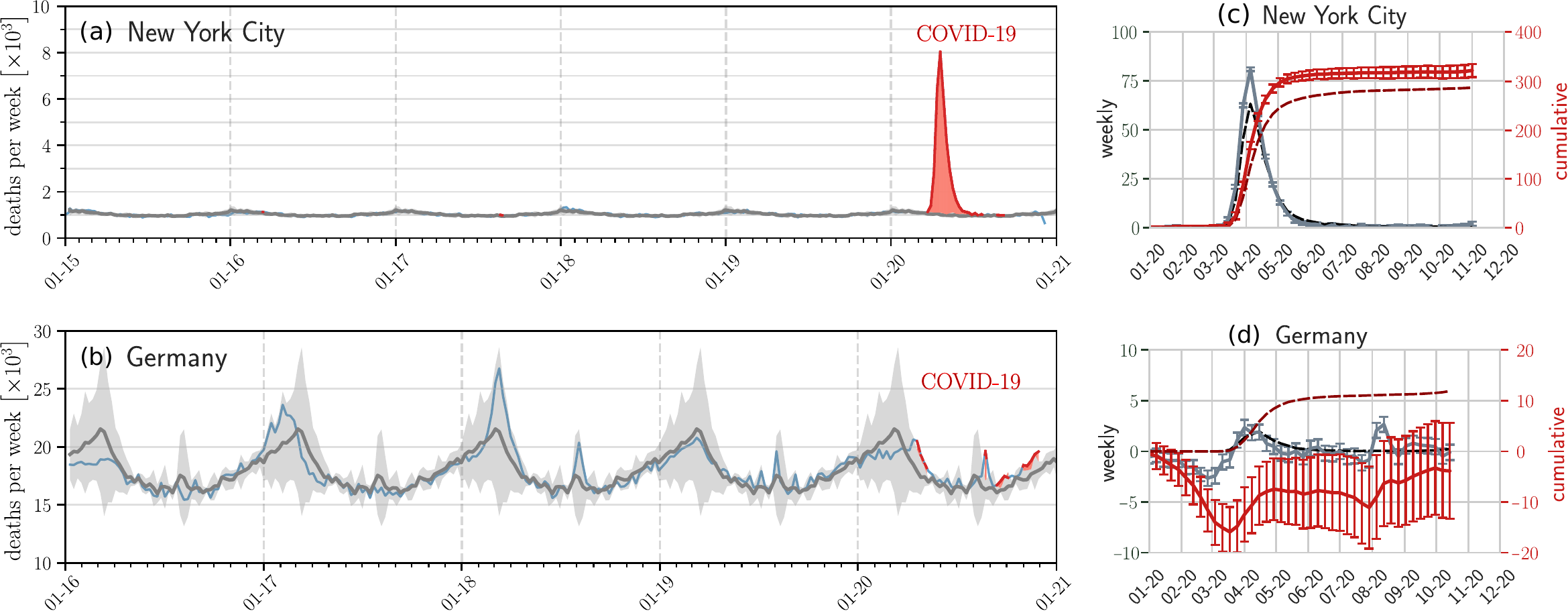}
\caption{\textbf{Examples of seasonal mortality and excess
    deaths}. The evolution of weekly deaths in (a) New York City (six
  years) and (b) Germany (five years) derived from data in
  Refs.~\cite{excessdata,ExcessDeathsEconomist}.  Grey solid lines and shaded
  regions represent the historical numbers of deaths and corresponding
  confidence intervals defined in Eq.~\eqref{DK_STATS}.  Blue solid
  lines indicate weekly deaths, and weekly deaths that lie outside the
  confidence intervals are indicated by solid red lines. The red
  shaded regions represent statistically significant mean cumulative
  excess deaths $D_{\rm e}$.  The reported weekly confirmed deaths
  $d^{(0)}_{\rm c}(i)$ (dashed black curves), reported cumulative
  confirmed deaths $D_{\rm c}(k)$ (dashed dark red curves), weekly
  excess deaths $\bar{d}_{\rm e}(i)$ (solid grey curves), and
  cumulative excess deaths $\bar{D}_{\rm e}(k)$ (solid red curves) are
  plotted in units of per 100,000 in (c) and (d) for NYC and Germany,
  respectively. The excess deaths and the associated 95\% confidence
  intervals given by the error bars are constructed from historical
  death data in (a-b) and defined in Eqs.~\eqref{DK_STATS} and
  \eqref{DE_STATS}. In NYC there is clearly a significant number of
  excess deaths that can be safely attributed to COVID-19, while to date
  in Germany, there have been no significant excess deaths.  Excess
  death data from other jurisdictions are shown in the Supplementary
  Information and typically show excess deaths greater than reported
  confirmed deaths (with Germany an exception as shown in (d)).}
%
\label{fig:panels}
\end{figure*}
In practice, deaths are typically tallied daily,
weekly~\cite{euromomo,USData}, or sometimes aggregated
monthly~\cite{ExcessDeathsEconomist,ExcessDeathsCDC} with historical
records dating back $J$ years so that for every period $i$ there are a
total of $J+1$ death values. We denote by $d^{(j)}(i)$ the total
number of deaths recorded in period $i$ from the $j^{\rm th}$ previous
year where $0 \leq j \leq J$ and where $j=0$ indicates the current
year.  In this notation, $D = D_{\rm c} + D_{\rm u} = \sum_i
d^{(0)}(i)$, where the summation tallies deaths over several periods
of interest within the pandemic. Note that we can decompose
$d^{(0)}(i) = d_{\rm c}^{(0)}(i) + d_{\rm u}^{(0)}(i)$, to include the
contribution from the confirmed and unconfirmed deaths during each
period $i$, respectively.
%
%
To quantify the total cumulative excess deaths we derive excess deaths
$d_{\rm e}^{(j)}(i)=d^{(0)}(i)-d^{(j)}(i)$ per week relative to the
$j^{\rm th}$ previous year.  Since $d^{(0)}(i)$ is the total number of
deaths in week $i$ of the current year, by definition $d_{\rm
  e}^{(0)}(i) \equiv 0$. The excess deaths during week $i$,
$\bar{d}_{\rm e}(i)$, averaged over $J$ past years and the associated,
unbiased variance $\sigma_{\rm e}(i)$ are given by

\begin{align}
\bar{d}_{\rm e}(i) & = {1\over J}\sum_{j=1}^{J} d_{\rm e}^{(j)}(i), \nonumber \\
\sigma_{{\rm e}}^{2}(i)& = {1\over J-1}\sum_{j=1}^{J}
\left[d_{\rm e}^{(j)}(i)-\bar{d}_{\rm e}(i)\right]^{2}.
\label{DK_STATS}
\end{align}
The corresponding quantities accumulated over $k$ weeks 
%
%
define the mean and variance of the cumulative excess deaths
$\bar{D}_{\rm e}(k)$ and $\Sigma_{{\rm e}}(k)$

\begin{align}
\bar{D}_{\rm e}(k) & = {1\over J}\sum_{j=1}^{J} \sum_{i=1}^k d_{\rm e}^{(j)}(i), \nonumber \\
\Sigma_{{\rm e}}^{2}(k) & = {1\over J-1}\sum_{j=1}^{J}
\left[\sum_{i=1}^k d_{\rm e}^{(j)}(i)-\bar{D}_{\rm e}(k) \right]^{2},
\label{DE_STATS}
\end{align}
where deaths are accumulated from the first to the $k^{\rm th}$ week
of the pandemic.  The variance in Eqs.~\eqref{DK_STATS} and
\eqref{DE_STATS} arise from the variability in the baseline number of
deaths from the same time period in $J$ previous years.

We gathered excess death statistics from over 23 countries and all US
states. Some of the data derive from open-source online repositories
as listed by official statistical bureaus and health ministries
\cite{USData,SpainData,EnglandData,SwitzerlandData,Istat,ExcessDeathsCDC};
other data are elaborated and tabulated in
Ref.~\cite{ExcessDeathsEconomist}. In some countries excess death
statistics are available only for a limited number of states or
jurisdictions (\textit{e.g.}, Brazil).
%
%
The US death statistics that we use in this study is based on weekly
death data between 2015--2019~\cite{ExcessDeathsCDC}. For all other
countries, the data collection periods are summarized in
Ref.~\cite{ExcessDeathsEconomist}. Fig.~\ref{fig:panels}(a-b) shows
historical death data for NYC and Germany, while
Fig.~\ref{fig:panels}(c-d) plots the confirmed and excess deaths and
their confidence levels computed from Eqs.~\eqref{DK_STATS} and
\eqref{DE_STATS}.  We assumed that the cumulative summation is performed
from the start of 2020 to the current week $k=K$ so that $\bar{D}_{\rm
  e}(K) \equiv \bar{D}_{\rm e}$ indicates excess deaths at the time of
writing. Significant numbers of excess deaths are clearly evident for
NYC, while Germany thus far has not experienced significant excess
deaths.

To evaluate CFR and $M$, data on only $D_{\rm c}, N_{\rm c}$, and
$R_{\rm c}$ are required, which are are tabulated by many
jurisdictions.  To estimate the numerators of IFR and ${\cal M}$, we
approximate $D_{\rm c}+D_{\rm u} \approx \bar{D}_{\rm e}$ using
Eq.~\eqref{DE_STATS}.
%
%
For the denominators, estimates of the unconfirmed infected $N_{\rm
  u}$ and unconfirmed recovered populations $R_{\rm u}$ are
required. In the next two sections we propose methods to estimate
$N_{\rm u}$ using a statistical testing model and $R_{\rm u}$ using
compartmental population model.

\subsection*{Statistical testing model with bias and testing errors}

The total number of confirmed and unconfirmed infected individuals
$N_{\rm c} + N_{\rm u}$ appears in the denominator of the IFR. To
better estimate the infected population we present a statistical model
for testing in the presence of bias in administration and testing
errors.  Although $N_{\rm c} + N_{\rm u}$ used to estimate the IFR
includes those who have died, depending on the type of test, it may or
may not include those who have recovered.  If $S, I, R, D$ are the
numbers of susceptible, currently infected, recovered, and deceased
individuals, the total population is $N =S + I + R + D$ and the
infected fraction can be defined as $f = (N_{\rm c} + N_{\rm u})/N=(I
+ R + D)/N$ for tests that include recovered and deceased individuals
(\textit{e.g.}, antibody tests), or $f = (N_{\rm c} + N_{\rm
  u})/N=(I+D)/N$ for tests that only count currently infected
individuals (\textit{e.g.}, RT-PCR tests). If we assume that the total
population $N$ can be inferred from census estimates, the problem of
identifying the number of unconfirmed infected persons $N_{\rm u}$ is
mapped onto the problem of identifying the true fraction $f$ of the
population that has been infected.
\begin{figure*}
\includegraphics[width=0.75\textwidth]{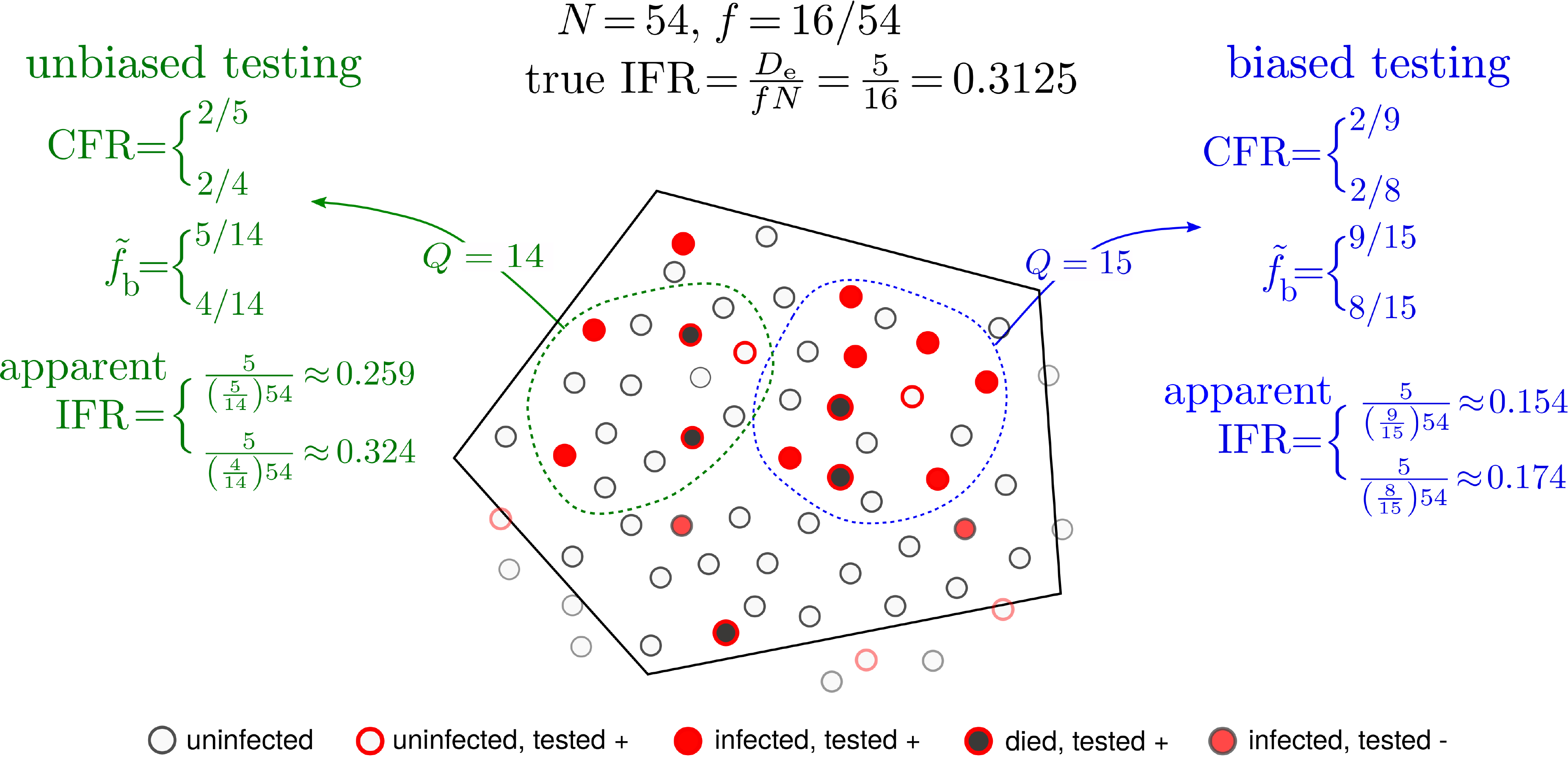}
\caption{\textbf{Biased and unbiased testing of a population.}
A hypothetical scenario of testing a population (total $N=54$
individuals) within a jurisdiction (solid black boundary). Filled red
circles represent the true number of infected individuals who tested
positive and the black-filled red circles indicate individuals who
have died from the infection. Open red circles denote uninfected
individuals who were tested positive (false positives) while filled
red circles with dark gray borders are infected individuals who were
tested negative (false negatives).  In the jurisdiction of interest 5
have died of the infection while 16 are truly infected. The true
fraction $f$ of infected in the entire population is thus $f=16/54$
and the true IFR=$5/16$.  However, under testing (green and blue)
samples, a false positive is shown to arise.  If the apparent positive
fraction $\tilde{f}_{\rm b}$ is derived from a biased sample (blue),
the estimated \textit{apparent} IFR can be quite different from the
true one.  For a less biased (more random) testing sample (green
sample), a more accurate estimate of the total number of infected
individuals is $N_{\rm c}+N_{\rm u} = \tilde{f}_{\rm b} N =
(5/14)\times 54 \approx 19$ when the single false positive in this
sample is included, and $\tilde{f}_{\rm b} N = (4/14)\times 54 \approx
15$ when the false positive is excluded, and allows us to more
accurately infer the IFR. Note that CFR is defined according to the
tested quantities $D_{\rm c}/N_{\rm c}$ which are precisely $2/9$ and
$2/5$ for the blue and green sample, respectively, if false positives
are considered. When false negatives are known and factored out
$\mathrm{CFR}=2/8$ and 2/4, for the blue and green samples,
respectively.}
\label{fig:metrics}
\end{figure*}

Typically, $f$ is determined by testing a representative sample and
measuring the proportion of infected persons within the
sample. Besides the statistics of sampling, two main sources of
systematic errors arise: the non-random selection of individuals to be
tested and errors intrinsic to the tests themselves. Biased sampling
arises when testing policies focus on symptomatic or at-risk
individuals, leading to over-representation of infected individuals.

Figure~\ref{fig:metrics} shows a schematic of a hypothetical initial
total population of $N=54$ individuals in a specified
jurisdiction. Without loss of generality we assume there are no
unconfirmed deaths, $D_{\rm u} = 0$, and that all confirmed deaths are
equivalent to excess deaths, so that $\bar{D}_{\rm e} = D_{\rm c} = 5$
in the jurisdiction represented by Fig.~\ref{fig:metrics}.  Apart from
the number of deceased, we also show the number of infected and
uninfected subpopulations and label them as true positives, false
positives, and false negatives. The true number of infected
individuals is $N_{\rm c} + N_{\rm u} = 16$ which yields the true $f =
16/54 = 0.27$ and an IFR = 5/16 = 0.312 within the jurisdiction.

Also shown in Fig.~\ref{fig:metrics} are two examples of sampling.
Biased sampling and testing is depicted by the blue contour in which 6
of the 15 are alive and infected, 2 are deceased, and the remaining 7
are healthy.  For simplicity, we start by assuming no testing errors.
This measured infected fraction of this sample $8/15 = 0.533 > f =
0.296$ is biased since it includes a higher proportion of infected
persons, both alive and deceased, than that of the entire
jurisdiction.  Using this biased measured infected fraction of $8/15$
yields ${\rm IFR} = 5/(0.533 \cdot 54) \approx 0.174$, which
significantly underestimates the true ${\rm IFR} = 0.312$.  A
relatively unbiased sample, shown by the green contour, yields an
infected fraction of $4/14 \approx 0.286$ and an apparent ${\rm
  IFR}\approx 0.324$ which are much closer to the true fraction $f$
and IFR. In both samples discussed above we neglected testing errors
such as false positives indicated in Fig.~\ref{fig:metrics}.  Tests
that are unable to distinguish false positives as negatives would
yield a larger $N_{\rm c}$, resulting in an apparent infected fraction
$9/15$ and an even smaller apparent ${\rm IFR}\approx 0.154$.  By
contrast, the false positive testing errors on the green sample would
yield an apparent infected fraction $5/15 = 0.333$ and IFR= 0.259.

Given that test administration can be biased, we propose a parametric
form for the apparent or measured infected fraction

\begin{equation}
f_{\rm b}=f_{\rm b}(f, b)\equiv {fe^{b} \over f(e^{b}-1)+1},
\label{F_B}
\end{equation}
to connect the apparent (biased sampling) infected fraction $f_{\rm
  b}$ with the true underlying infection fraction.  The bias parameter
$-\infty < b < \infty$ describes how an infected or uninfected
individual might be preferentially selected for testing, with $b <0$
(and $f_{\rm b}< f$) indicating under-testing of infected individuals,
and $b>0$ (and $f_{\rm b}> f$) representing over-testing of infecteds.
%
%
A truly random, unbiased sampling arises only when $b=0$ where $f_{\rm
  b} = f$. 
%
%
Given $Q$ (possibly biased) tests to date, testing errors, and
ground-truth infected fraction $f$, we derive in the SI the likelihood
of observing a positive fraction $\tilde{f}_{\rm b} = \tilde{Q}^{+}/Q$
(where $\tilde{Q}^{+}$ is the number of recorded positive tests):

\begin{equation}
P(\tilde{f}_{\rm b}\vert \theta)\approx
{1\over  \sqrt{2\pi}\sigma_{\rm T}}
\exp\left[-{(\tilde{f}_{\rm b} -\mu)^{2}\over 
2\sigma_{\rm T}^{2}}\right],
\label{PTOT}
\end{equation}
in which

\begin{eqnarray}
\mu &\equiv &
f_{\rm b}(f,b)(1-{\rm FNR}) + (1-f_{\rm b}(f,b)){\rm FPR}, 
\nonumber \\
\sigma_{\rm T}^{2} 
&\equiv & \mu (1-\mu)/Q.
\label{MU_SIGMA}
\end{eqnarray}
Here, $\mu$ is the expected value of the measured and biased fraction
$\tilde{f}_{\rm b}$ and $\sigma_{\rm T}^{2}$ is its variance.  Note
that the parameters $\theta =\{Q, f, b, {\rm FPR}, {\rm FNR}\}$ may be
time-dependent and change from sample to sample.  Along with the
likelihood function $P(\tilde{f}_{\rm b}\vert f,\theta)$, one can also
propose a prior distribution $P(\theta\vert \alpha)$ with
hyperparameters $\alpha$, and apply Bayesian methods to infer $\theta$
(see SI).

To evaluate IFR, we must now estimate $f$ given $\tilde{f}_{\rm b} =
\tilde{Q}^{+}/Q$ and possible values for ${\rm FPR}$, ${\rm FNR}$,
and/or $b$, or the hyperparameters $\alpha$ defining their
uncertainty. The simplest maximum likelihood estimate of $f$ can be
found by maximizing $P(\tilde{f}_{\rm b}\vert \theta)$ with respect to
$f$ given a measured value $\tilde{f}_{\rm b}$ and all other parameter
values $\theta$ specified:
\begin{equation}
\hat{f} \approx {\tilde{f}_{\rm b} -{\rm FPR}
\over e^{b}(1-{\rm FNR}-\tilde{f}_{\rm b})+ \tilde{f}_{\rm b}-{\rm FPR}}.
\label{f_b_hat}
\end{equation}
Note that although FNRs are typically larger than FPRs, small values
of $f$ and $\tilde{f}_{\rm b}$ imply that $\hat{f}$ and $\mu$ are more
sensitive to the FPR, as indicated by Eqs.~\eqref{MU_SIGMA} and
\eqref{f_b_hat}.

If time series data for $\tilde{f}_{\rm b}= \tilde{Q}^{+}/Q$ are
available, one can evaluate the corrected testing fractions in 
Eq.\,\eqref{f_b_hat} for each time interval. Assuming that serological
tests can identify infected individuals long after symptom onset, the
latest value of $\hat{f}$ would suffice to estimate corresponding
mortality metrics such as the $\mathrm{IFR}$.  For RT-PCR testing, one
generally needs to track how $\tilde{f}_{\rm b}$ evolves in time. A
rough estimate would be to use the mean of $\tilde{f}_{\rm b}$ over
the whole pandemic period to provide a lower bound of the estimated
prevalence $\hat{f}$.

The measured $\tilde{f}_{\rm b}$ yields only the apparent ${\rm IFR} =
\bar{D}_{\rm e}/(\tilde{f}_{\rm b} N)$, but Eq.~\eqref{f_b_hat} can
then be used to evaluate the corrected ${\rm IFR} \approx \bar{D}_{\rm
  e}/(\hat{f} N)$ which will be a better estimate of the true IFR.
For example, under moderate bias $\vert b\vert \lesssim 1$ and
assuming FNR, FPR, $\tilde {f}_{\rm {b}} \lesssim 1$
Eq.~\eqref{f_b_hat} relates the apparent and corrected IFRs through
$\bar{D}_{\rm e}/(\tilde{f}_{\rm b} N)\sim \bar{D}_{\rm e}/((\hat{f}
e^{\rm b} + {\rm{FPR}}) N)$.

Another commonly used representation of the IFR is $\mathrm{IFR} = p
(D_{\rm c}+D_{\rm u})/N_{\rm c} = p\bar{D}_{\rm e}/N_{\rm c}$. This
expression is equivalent to our $\mathrm{IFR} = \bar{D}_{\rm e}/(f N)$
if $p = N_{\rm c}/(N_{\rm c}+ N_{\rm u})\approx \tilde{Q}^{+}/(fN)$ is
defined as the fraction of infected individuals that are
confirmed~\cite{LI_SCIENCE,chow2020global}. In this alternative
representation, the $p$ factor implicitly contains the effects of
biased testing. Our approach allows the true infected fraction $f$ to
be directly estimated from $\tilde{Q}^{+}$ and $N$.

While the estimate $\hat{f}$ depends strongly on $b$ and ${\rm FPR}$,
and weakly on ${\rm FNR}$, the \textit{uncertainty} in $f$ will depend
on the uncertainty in the values of $b$, ${\rm FPR}$, and ${\rm FNR}$.
A Bayesian framework is presented in the SI, but under a a Gaussian
approximation for all distributions, the uncertainty in the testing
parameters can be propagated to the squared coeffcient
$\sigma_{f}^{2}/\hat{f}^{2}$ of variation of the estimated infected
fraction $\hat{f}$, as explicitly computed in the SI.  Moreover, the
uncertainties in the mortality indices $Z$ decomposed into the
uncertainties of their individual components are listed in
Table~\ref{tab:mortality_error}.
\subsection*{Using compartmental models to estimate resolved mortalities}
Since the number of unreported recovered individuals $R_{\rm u}$
required to calculate ${\cal M}$ is not directly related to excess
deaths nor to positive-tested populations, we use an SIR-type
compartmental model to relate $R_{\rm u}$ to other inferable
quantities~\cite{keeling2011modeling}.  Both unconfirmed recovered
individuals and unconfirmed deaths are related to unconfirmed infected
individuals who recover at rate $\gamma_{\rm u}$ and die at rate
$\mu_{\rm u}$. The equations for the cumulative numbers of unconfirmed
recovered individuals and unconfirmed deaths,

\begin{equation}
\frac{\dd R_{\rm u}(t)}{\dd t} =  \gamma_{\rm u}(t) I_{\rm u}(t),\qquad
\frac {\dd D_{\rm u}(t)}{\dd t} = \mu_{\rm u}(t) I_{\rm u}(t),
\label{eq:SIR_ODE}
\end{equation}
can be directly integrated to find $R_{\rm u}(t) = \int_{0}^{t}
\gamma_{\rm u}(t')I_{\rm u}(t')\dd t'$ and $D_{\rm u}(t) =
\int_{0}^{t} \mu_{\rm u}(t')I_{\rm u}(t')\dd t'$.  The rates
$\gamma_{\rm u}$ and $\mu_{\rm u}$ may differ from those averaged over
the entire population since testing may be biased towards
subpopulations with different values of $\gamma_{\rm u}$ and $\mu_{\rm
  u}$.  If one assumes $\gamma_{\rm u}$ and $\mu_{\rm u}$ are
approximately constant over the period of interest, we find $R_{\rm
  u}/D_{\rm u} \approx \gamma_{\rm u}/\mu_{\rm u}\equiv \gamma$.  We
now use $D_{\rm u} = \bar{D}_{\rm e} - D_{\rm c}$, where both
$\bar{D}_{\rm e}$ and $D_{\rm c}$ are given by data, to estimate
$R_{\rm u} \approx \gamma (\bar{D}_{\rm e} - D_{\rm c})$ and write
$\mathcal{M}$ as
\begin{equation}
\mathcal{M} =\frac{\displaystyle{\bar{D}_{\rm e}}} 
{\bar{D}_{\rm e} + R_{\rm c} + \gamma
(\bar{D}_{\rm e}-D_{\rm c})}.
\label{eq:Mp_2}
\end{equation}
Thus, a simple SIR model transforms the problem of determining the
number of unreported death and recovered cases in $\mathcal{M}$ to the
problem of identifying the recovery and death rates in the untested
population. Alternatively, we can make use of the fact that both the
IFR and resolved mortality $\mathcal{M}$ should have comparable values
and match ${\cal M}$ to $\mathrm{IFR}\approx
0.1-1.5$\%~\cite{salje2020estimating,chow2020global,ioannidis2020infection}
by setting $\gamma \equiv \gamma_{\rm u}/\mu_{\rm u}\approx 100-1000$
(see SI for further information). Note that inaccuracies in confirming
deaths may give rise to $D_{\rm c} > \bar{D}_{\rm e}$. Since by
definition, infection-caused excess deaths must be greater than the
confirmed deaths, we set $\bar{D}_{\rm e}-D_{\rm c} = 0$ whenever data
happens to indicate $\bar{D}_{\rm e}$ to be less than $D_{\rm c}$.

\section*{Results}

Here, we present much of the available worldwide fatality data,
construct the excess death statistics, and compute mortalities and
compare them across jurisdictions. We show that standard mortality
measures significantly underestimate the death toll of COVID-19 for
most regions (see Figs.~\ref{fig:panels} and
\ref{fig:excess_rate}). We also use the data to estimate uncertainties
in the mortality measures and relate them uncertainties of the
underlying components and model parameters.
\subsection*{Excess and confirmed deaths} 
%
%
%

We find that in New York City for example, the number of confirmed
COVID-19 deaths between March 10, 2020 and December 10, 2020 is
19,694~\cite{NYCdeaths} and thus significantly lower than the 27,938
(95\% CI 26,516--29,360) reported excess mortality
cases~\cite{USData}. From March 25, 2020 until December 10, 2020,
Spain counts 65,673 (99\% confidence interval [CI] 91,816--37,061)
excess deaths~\cite{SpainData}, a number that is substantially larger
than the officially reported 47,019 COVID-19
deaths~\cite{AllReps}. The large difference between excess deaths and
reported COVID-19 deaths in Spain and New York City is also observed
in Lombardia, one of the most affected regions in Italy. From February
23, 2020 until April 4, 2020, Lombardia reported 8,656 reported
COVID-19 deaths~\cite{AllReps} but 13,003 (95\% 12,335--13,673) excess
deaths~\cite{Istat}. Starting April 5 2020, mortality data in
Lombardia stopped being reported in a weekly format.  In
England/Wales, the number of excess deaths from the onset of the
COVID-19 outbreak on March 1, 2020 until November 27, 2020 is 70,563
(95\% CI 52,250--88,877) whereas the number of reported COVID-19
deaths in the same time interval is 66,197~\cite{excessdata}. In
Switzerland, the number of excess deaths from March 1, 2020 until
November 29, 2020 is 5,664 (95\% CI
4,281--7,047)~\cite{SwitzerlandData}, slightly larger than the
corresponding 4,932 reported COVID-19 deaths~\cite{AllReps}.

\begin{figure}
\includegraphics{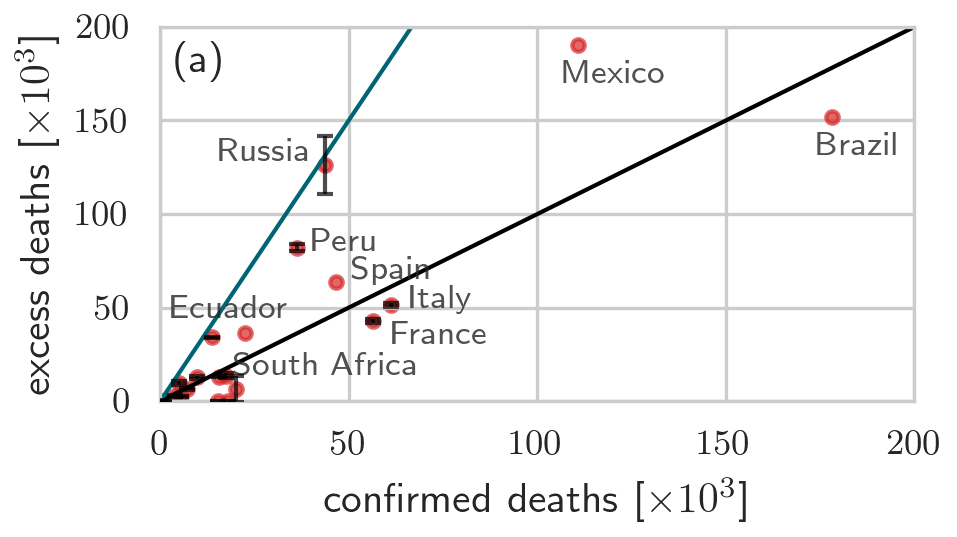}
\includegraphics{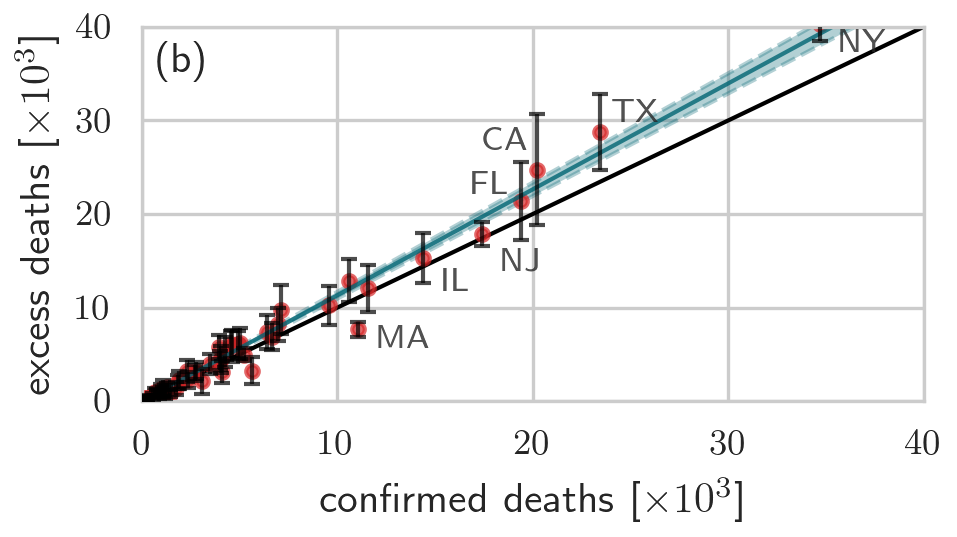}
\caption{\textbf{Excess deaths versus confirmed deaths across
    different countries/states.} The number of excess deaths in 2020
  versus confirmed deaths across different countries (a) and US states
  (b). The black solid lines in both panels have slope 1. In (a) the
  blue solid line is a guide-line with slope 3; in (b) the blue solid
  line is a least-squares fit of the data with slope $1.132$ (95\% CI
  1.096--1.168; blue shaded region). All data were updated on December
  10,
  2020~\cite{ExcessDeathsEconomist,ExcessDeathsCDC,NYCData,dong2020interactive}.}
\label{fig:excess_versus_confirmed} 
\end{figure}

To illustrate the significant differences between excess deaths and
reported COVID-19 deaths in various jurisdictions, we plot the excess
deaths against confirmed deaths for various countries and US states as
of December 10, 2020 in Fig.~\ref{fig:excess_versus_confirmed}. We
observe in Fig.~\ref{fig:excess_versus_confirmed}(a) that the number
of excess deaths in countries like Mexico, Russia, Spain, Peru, and
Ecuador is significantly larger than the corresponding number of
confirmed COVID-19 deaths. In particular, in Russia, Ecuador, and
Spain the number of excess deaths is about three times larger than the
number of reported COVID-19 deaths. As described in the Methods
section, for certain countries (\eg, Brazil) excess death data is not
available for all states~\cite{ExcessDeathsEconomist}. For the
majority of US states the number of excess deaths is also larger than
the number of reported COVID-19 deaths, as shown in
Fig.~\ref{fig:excess_versus_confirmed}(b).  We performed a
least-square fit to calculate the proportionality factor $m$ arising
in $\bar{D}_{\rm e} = m D_{\rm c}$ and found $m \approx 1.132$ (95\%
CI 1.096--1.168). That is, across all US states, the number of excess
deaths is about 13\% larger than the number of confirmed COVID-19
deaths.

\subsection*{Estimation of mortality measures and their uncertainties}

We now use excess death data and the statistical and modeling
procedures to estimate mortality measures $Z=$ IFR, CFR, $M$, ${\cal
  M}$ across different jurisdictions, including all US states and more
than two dozen countries.\footnote{We provide an online dashboard that
  shows the real-time evolution of CFR and $M$ at
  \url{https://submit.epidemicdatathon.com/\#/dashboard}}.
%
%
%
Accurate estimates of the confirmed $N_{\rm c}$ and dead $D_{\rm c}$
infected are needed to evaluate the CFR.  Values for the parameters
$Q$, FPR, FNR, and $b$ are needed to estimate $N_{\rm c}+N_{\rm u} = f
N$ in the denominator of the IFR, while $\bar{D}_{\rm e}$ is needed to
estimate the number of infection-caused deaths $D_{\rm c}+D_{\rm u}$
that appear in the numerator of the IFR and ${\cal M}$. Finally, since
we evaluate the resolved mortality $\mathcal{M}$, through
Eq.\,\ref{eq:Mp_2}, estimates of $\bar{D}_{\rm e}, D_{\rm c}, R_{\rm
  c}$, $\gamma$, and FPR, FNR (to correct for testing inaccuracies in
$D_{\rm c}$ and $R_{\rm c}$) are necessary. Whenever uncertainties are
available or inferable from data, we also include them in our
analyses.

Estimates of excess deaths and infected populations themselves suffer
from uncertainty encoded in the variances $\Sigma_{{\rm e}}^{2}$ and
$\sigma_{f}^{2}$. These uncertainties depend on uncertainties arising
from finite sampling sizes, uncertainty in bias $b$ and uncertainty in
test sensitivity and specificity, which are denoted $\sigma_{b}^{2}$,
$\sigma_{\rm I}^{2}$, and $\sigma_{\rm II}^{2}$, respectively. We use
$\Sigma^2$ to denote population variances and $\sigma^2$ to denote
parameter variances; covariances with respect to any two variables
${X,Y}$ are denoted as $\Sigma_{X,Y}$. Variances in the confirmed
populations are denoted $\Sigma_{N_{\rm c}}^{2}$, $\Sigma_{R_{\rm
    c}}^{2}$, and $\Sigma_{D_{\rm c}}^{2}$ and also depend on
uncertainties in testing parameters $\sigma_{\rm I}^{2}$ and
$\sigma_{\rm II}^{2}$.  The most general approach would be to define a
probability distribution or likelihood for observing some value of the
mortality index in $[Z, Z+\dd Z]$.  As outlined in the SI, these
probabilities can depend on the mean and variances of the components
of the mortalities, which in turn may depend on hyperparameters that
determine these means and variances.  Here, we simply assume
uncertainties that are propagated to the mortality indices through
variances in the model parameters and
hyperparameters~\cite{lee2006analyzing}. The squared coefficients of
variation of the mortalities are found by linearizing them about the
mean values of the underlying components and are listed in
Table~\ref{tab:mortality_error}.

\begin{table*}[htb]
\renewcommand*{\arraystretch}{2.8}
\begin{tabular}{|c|*{3}{c|}}\hline \makebox[4em]{Mortality $Z$} & \makebox[6em]{Uncertainties} & \makebox[8em]{CV$^{2}\displaystyle ={\Sigma_{Z}^{2}\over Z^{2}}$}
 \\\hline\hline $\displaystyle {\rm CFR}={D_{\rm c}\over N_{\rm c}}$
 \,\, & $\Sigma_{D_{\rm c}}^{2}$, $\Sigma_{N_{\rm c}}^{2}$, $\Sigma_{D_{\rm c}, N_{\rm c}}$ & 
$\displaystyle {\Sigma_{D_{\rm c}}^{2} \over D_{\rm c}^{2}}+
{\Sigma_{N_{\rm c}}^{2} \over N_{\rm c}^{2}} - 2{\Sigma_{D_{\rm c}, N_{\rm c}}
\over D_{\rm c}N_{\rm c}}$ \\[3pt]\hline
$\displaystyle {\rm IFR}={D_{\rm c} + D_{\rm u}\over N_{\rm c} +
  N_{\rm u}}\approx {\bar{D}_{\rm e} \over f N}$ & $\Sigma_{\rm e}^{2},
\Sigma_{N}^{2}, \Sigma_{{\rm e},N}, \sigma_{f}^{2}$ & $\displaystyle {{\sigma}_{f}^{2} \over
  \hat{f}^{2}}+ {\Sigma_{\rm e}^{2} \over \bar{D}_{\rm
    e}^{2}}+{\Sigma_{N}^{2} \over N^{2}}- {2\Sigma_{{\rm e}, N}\over
  \bar{D}_{\rm e}N}$ 
\\[3pt]\hline
$\displaystyle M={D_{\rm c}\over D_{\rm c}+R_{\rm c}}$ & 
$\Sigma_{D_{\rm c}}^{2}, \Sigma_{R_{\rm c}}^{2}, \Sigma_{D_{\rm c}, R_{\rm c}}$ & 
$\displaystyle M^{2}\left({R_{\rm c}\over D_{\rm c}}\right)^{2}
\left[{\Sigma_{D_{\rm c}}^{2} \over D_{\rm c}^{2}}+ {\Sigma_{R_{\rm
        c}}^{2} \over R_{\rm c}^{2}} - {2\Sigma_{R_{\rm c}, D_{\rm
        c}}\over R_{\rm c}D_{\rm c}}\right]$
\\[3pt]\hline
$\displaystyle {\cal M}={\bar{D}_{\rm e} \over \bar{D}_{\rm e} +
  R_{\rm c} +R_{\rm u}}$ & $\Sigma_{\rm e}^{2}, \Sigma_{R_{\rm c}}^{2}, 
\Sigma_{R_{\rm u}}^{2}, \Sigma_{R_{\rm c}, R_{\rm u}}$ & 
$\displaystyle (1-{\cal M})^{2}{\Sigma_{\rm e}^{2} \over \bar{D}_{\rm e}^{2}} 
+{\Sigma_{R_{\rm c}}^{2}\over \Gamma^{2}}
+{\Sigma_{R_{\rm u}}^{2}\over \Gamma^{2}} 
-{2\Sigma_{R_{\rm c}, R_{\rm u}}\over \Gamma^{2}}$
\\[3pt]\hline
$\displaystyle {\cal M}={\bar{D}_{\rm e} \over 
\bar{D}_{\rm e} + R_{\rm c} + \gamma(\bar{D}_{\rm e}-D_{\rm c})}$ & 
$\Sigma_{R_{\rm c}}^{2}, \Sigma_{\rm e}^{2}, \Sigma_{R_{\rm c}, \gamma},
\sigma_{\gamma}^{2}$ & $\displaystyle (1-{\cal M})^{2}
{\Sigma_{\rm e}^{2} \over \bar{D}_{\rm e}^{2}} + 
{\Sigma_{R_{\rm c}}^{2} \over \Gamma^{2}} + 
{(\bar{D}_{\rm e}-D_{\rm c})^{2}\sigma_{\gamma}^{2} \over \Gamma^{2}} -
{2(\bar{D}_{\rm e}-D_{\rm c})\Sigma_{R_{\rm c}, \gamma}\over \Gamma^{2}}$ \\[3pt] \hline
\end{tabular}
\caption{\textbf{Uncertainty propagation for different mortality
    measures.}  Table of squared coefficients of variation
  $\mathrm{CV}^2=\Sigma_{Z}^{2}/Z^{2}$ for the different mortality
  indices $Z$ derived using standard error propagation
  expansions~\cite{lee2006analyzing}. We use $\Sigma_{N}^{2},
  \Sigma_{N_{\rm c}}^2$, $\Sigma_{R_{\rm c}}^2$, and $\Sigma_{D_{\rm
      c}}^2$ to denote the uncertainties in the total population,
  confirmed cases, recoveries, and deaths, respectively.  The variance
  of the number of excess deaths is $\Sigma_{{\rm e}}^2$, which
  feature in the IFR and ${\cal M}$.  The uncertainty in the infected
  fraction $\sigma_{f}^{2}$ that contributes to the uncertainty in IFR
  depends on uncertainties in testing bias and testing errors as shown
  in Eq.~\eqref{sigma_f}. The term $\Sigma_{D_{\rm c}, N_{\rm c}}$
  represents the covariance between $D_{\rm c}, N_{\rm c}$, and
  similarly for all other covariances $\Sigma_{{\rm e}, N}$,
  $\Sigma_{D_{\rm c}, R_{\rm c}}$, $\Sigma_{R_{\rm c}, R_{\rm u}}$,
  $\Sigma_{R_{\rm c}, \gamma}$.  Since variations in $D_{\rm e}$ arise
  from fluctuations in past-year baselines and not from current
  intrinsic uncertainty, we can neglect correlations between
  variations in $D_{\rm e}$ and uncertainty in $R_{\rm c}, R_{\rm u}$.
  In the last two rows, representing ${\cal M}$ expressed in two
  different ways, $\Gamma \equiv \bar{D}_{\rm e} + R_{\rm c} + R_{\rm
    u}$ and $\bar{D}_{\rm e} + R_{\rm c} + \gamma (\bar{D}_{\rm e} -
  D_{\rm c})$, respectively. Moreover, when using the SIR model to
  replace $D_{\rm u}$ and $R_{\rm u}$ with $\bar{D}_{\rm e}-D_{\rm
    c}\geq 0$, there is no uncertainty associated with $D_{\rm u}$ and
  $R_{\rm u}$ in a deterministic model. Thus, covariances cannot be
  defined except through the uncertainty in the parameter $\gamma =
  \gamma_{\rm u}/\mu_{\rm u}$.}
\label{tab:mortality_error}
\end{table*}

\begin{figure}
\includegraphics{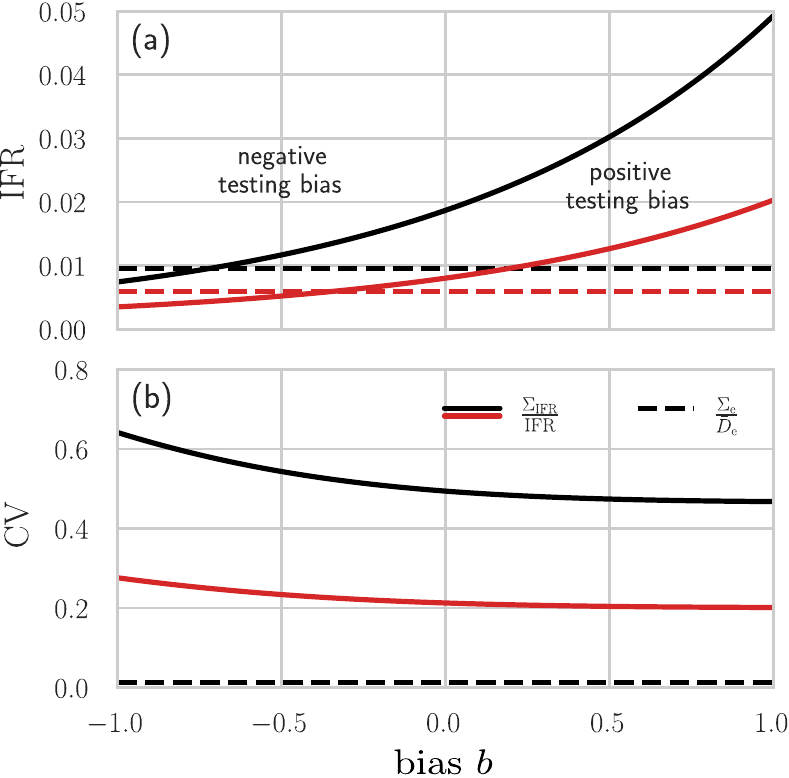}
\caption{\textbf{Different mortality measures across different
    regions.} (a) The apparent (dashed lines) and corrected (solid
  lines) IFR in the US, as of November 1, 2020, estimated using excess
  mortality data. We set $\tilde{f}_b=0.093,0.15$ (black,red), ${\rm
    FPR}=0.05$, ${\rm FNR}=0.2$, and $N=330$ million. For the
  corrected IFR, we use $\hat{f}$ as defined in
  Eq.~\eqref{f_b_hat}. Unbiased testing corresponds to setting
  $b=0$. For $b>0$ (positive testing bias), infected individuals are
  overrepresented in the sample population. Hence, the corrected
  $\mathrm{IFR}$ is larger than the apparent $\mathrm{IFR}$. If $b$ is
  sufficiently small (negative testing bias), the corrected
  $\mathrm{IFR}$ may be smaller than the apparent $\mathrm{IFR}$. (b)
  The coefficient of variation of $D_{\rm e}$ (dashed line) and IFR
  (solid lines) with $\sigma_{\rm I}=0.02$, $\sigma_{\rm II}=0.05$,
  and $\sigma_{b}=0.2$ (see Tab.~\ref{tab:mortality_error}).}
\label{fig:IFR} 
\end{figure}
To illustrate the influence of different biases $b$ on the
$\mathrm{IFR}$ we use $\hat{f}$ from Eq.~\eqref{f_b_hat} in the
corrected $\mathrm{IFR}\approx \bar{D}_{\rm e}/(\hat{f} N)$. We model
RT-PCR-certified COVID-19 deaths~\cite{CDC_classifying} by setting the
${\rm FPR}=0.05$~\cite{watson2020interpreting} and the ${\rm
  FNR}=0.2$~\cite{fang2020sensitivity,wang2020detection}. The
observed, possibly biased, fraction of positive tests $\tilde{f}_{\rm
  b}=\tilde{Q}^+/Q$ can be directly obtained from corresponding
empirical data.  As of November 1, 2020, the average of
$\tilde{f}_{\rm b}$ over all tests and across all US states is about
9.3\%~\cite{CDCPositivityRate}. The corresponding number of excess
deaths is $\bar{D}_{\rm{e}}=294,700$~\cite{ExcessDeathsEconomist} and
the US population is about $N\approx 330$ million~\cite{Census}. To
study the influence of variations in $\tilde{f}_{\rm b}$, in addition
to $\tilde{f}_{\rm b}=0.093$, we also use a slightly larger
$\tilde{f}_{\rm b}=0.15$ in our analysis. In Fig.~\ref{fig:IFR} we
show the apparent and corrected $\mathrm{IFR}$s for two values of
$\tilde{f}_{\rm b}$ [Fig.~\ref{fig:IFR}(a)] and the coefficient of
variation $\rm{CV}_{\rm IFR}$ [Fig.~\ref{fig:IFR}(b)] as a function of
the bias $b$ and as made explicit in
Table~\ref{tab:mortality_measures}.  For unbiased testing [$b=0$ in
  Fig.\,\ref{fig:IFR}(a)], the corrected $\mathrm{IFR}$ in the US is
1.9\% assuming $\tilde{f}_{\rm b}=0.093$ and 0.8\% assuming
$\tilde{f}_{\rm b}=0.15$. If $b>0$, there is a testing bias towards
the infected population, hence, the apparent $\mathrm{IFR} =
\bar{D}_{\rm e}/(\tilde{f}_{\rm b}N)$ is smaller than the corrected
$\mathrm{IFR}$ as can be seen by comparing the solid (corrected IFR)
and the dashed (apparent IFR) lines in Fig.~\ref{fig:IFR}(a).  For
testing biased towards the uninfected population ($b<0$), the
corrected $\mathrm{IFR}$ may be smaller than the apparent
$\mathrm{IFR}$. To illustrate how uncertainty in $\mathrm{FPR}$,
$\mathrm{FNR}$, and $b$ affect uncertainty in IFR, we evaluate
CV$_{\rm {IFR}}$ as given in Table~\ref{tab:mortality_error}.


The first term in uncertainty $\sigma_{f}^{2}/\hat{f}^{2}$ given in
Eq.~\eqref{sigma_f} is proportional to $1/Q$ and can be assumed to be
negligibly small, given the large number $Q$ of tests administered.
The other terms in Eq.\,\eqref{sigma_f} are evaluated by assuming
$\sigma_{b}=0.2, \sigma_{\rm I}=0.02$, and $\sigma_{\rm II}=0.05$ and
by keeping FPR = 0.05 and FNR = 0.2. Finally, we infer $\Sigma_{\rm
  e}$ from empirical data, neglect correlations between $D_{\rm e}$
and $N$, and assume that the variation in $N$ is negligible so that
$\Sigma_{{\rm e},N} = \Sigma_{N} \approx 0$.  Fig.~\ref{fig:IFR}(b)
plots ${\rm CV}_{\mathrm{IFR}}$ and ${\rm CV}_{D_{\rm e}}$ in the US
as a function of the underlying bias $b$. The coefficient of variation
${\rm CV}_{D_{\rm e}}$ is about 1\%, much smaller than ${\rm
  CV}_{\mathrm{IFR}}$, and independent of $b$.  For the values of $b$
shown in Fig.~\ref{fig:IFR}(b), ${\rm CV}_{\mathrm{IFR}}$ is between
47--64\% for $\tilde{f}_{\rm b}=0.093$ and between 20--27\% for
$\tilde{f}_{\rm b}=0.15$.

Next, we compared the mortality measures $Z=$IFR, CFR, $M$, $\cal M$
and the relative excess deaths $r$ listed in
Tab.~\ref{tab:mortality_measures} across numerous jurisdictions. To
determine the CFR, we use the COVID-19 data of
Refs.~\cite{NYCData,dong2020interactive}. For the apparent IFR, we use
the representation IFR $= p\bar{D}_{\rm e}/N_{\rm c}$ discussed
above. Although $p$ may depend on the stage of the pandemic, typical
estimates range from 4\%~\cite{hortaccsu2020estimating} to
10\%~\cite{chow2020global}.  We set $p=0.1$ over the lifetime of the
pandemic. We can also use the apparent IFR $= \bar{D}_{\rm e}/( f N)$,
however estimating the corrected IFR requires evaluating the bias $b$.
\begin{figure*}[htb]
\includegraphics{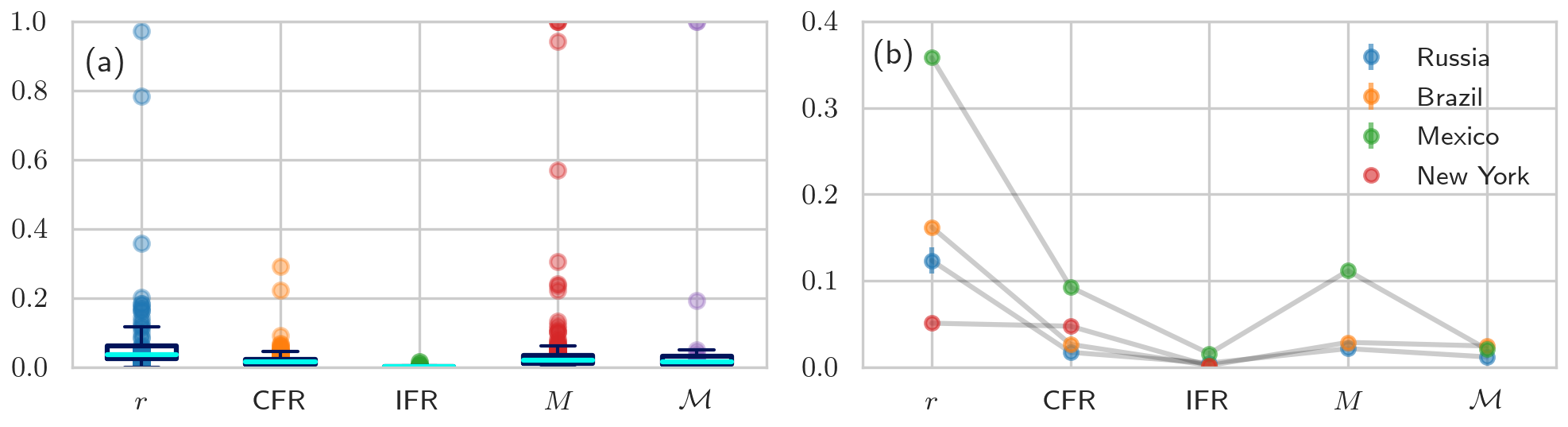}
\includegraphics{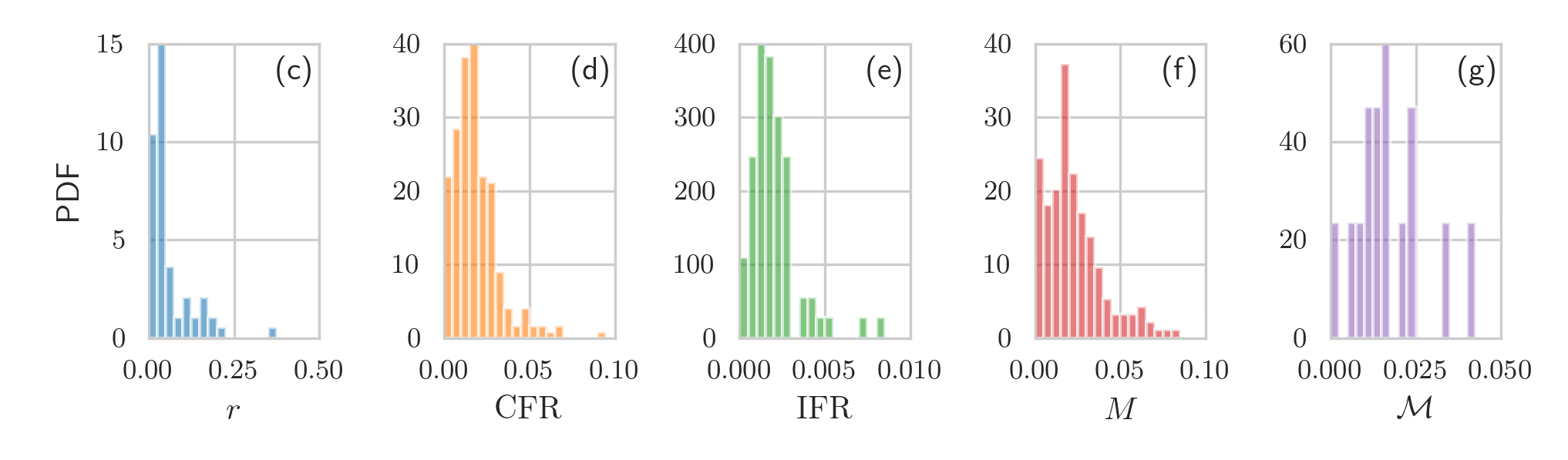}
\caption{\textbf{Mortality characteristics in different countries and
    states.} (a) The values of relative excess deaths $r$, the CFR,
  the $\mathrm{IFR}= p \bar{D}_{\mathrm{e}}/N_{\rm c}$ with $p=N_{\rm
    c}/(N_{\rm c}+N_{\rm u}) = 0.1$~\cite{chow2020global}, the
  confirmed resolved mortality $M$, and the true resolved mortality
  $\mathcal{M}$ (using $\gamma=100$) are plotted for various
  jurisdictions. (b) Different mortality measures provide ambiguous
  characterizations of disease severeness. (c--g) The probability
  density functions (PDFs) of the mortality measures shown in (a) and
  (b). Note that there are only very incomplete recovery data
  available for certain countries (\textit{e.g.}, US and UK). For
  countries without recovery data, we could not determine $M$ and
  $\mathcal{M}$. The number of jurisdictions that we used in (a) and (c--g)
  are 77, 246, 73, 191, and 21 for the respective mortality measures
  (from left to right). All data were updated December 10,
  2020~\cite{ExcessDeathsEconomist,ExcessDeathsCDC,NYCData,dong2020interactive}.}
\label{fig:boxplot} 
\end{figure*}
In Fig.~\ref{fig:boxplot}(a), we show the values of the relative
excess deaths $r$, the CFR, the apparent IFR, the confirmed resolved
mortality $M$, and the true resolved mortality $\mathcal{M}$ for
different (unlabeled) regions.  In all cases we set $p = 0.1,
\gamma=100$.  As illustrated in Fig.~\ref{fig:boxplot}(b), some
mortality measures suggest that COVID-induced fatalities are lower in
certain countries compared to others, whereas other measures indicate
the opposite.  For example, the total resolved mortality $\mathcal{M}$
for Brazil is larger than for Russia and Mexico, most likely due to
the relatively low number of reported excess deaths as can be seen
from Fig.~\ref{fig:excess_versus_confirmed} (a). On the other hand,
Brazil's values of $\mathrm{CFR}$, $\mathrm{IFR}$, and $M$ are
substantially smaller than those of Mexico [see
  Fig.~\ref{fig:boxplot}(b)].

The distributions of all measures $Z$ and relative excess deaths $r$
across jurisdictions are shown Fig.~\ref{fig:boxplot}(c--g) and encode
the global uncertainty of these indices. We also calculate the
corresponding mean values across jurisdictions, and use the empirical
cumulative distribution functions to determine confidence
intervals. The mean values across all jurisdictions are
$\widebar{r}=0.08$ (95\% CI 0.0025--0.7800),
$\widebar{\mathrm{CFR}}=0.020$ (95\% CI 0.0000--0.0565),
$\widebar{\mathrm{IFR}}=0.0024$ (95\% CI 0.0000--0.0150),
$\widebar{M}=0.038$ (95\% CI 0.0000--0.236), and
$\widebar{\mathcal{M}}=0.027$ (95\% CI 0.000--0.193). For calculating
$\widebar{M}$ and $\widebar{\mathcal{M}}$, we excluded countries with
incomplete recovery data. The distributions plotted in
Fig.~\ref{fig:boxplot}(c--g) can be used to inform our analyses of
uncertainty or heterogeneity as summarized in
Tab.~\ref{tab:mortality_error}.  For example, the overall variance
$\Sigma_{Z}^{2}$ can be determined by fitting the corresponding
empirical $Z$ distribution shown in Fig.~\ref{fig:boxplot}(c--g).
Table~\ref{tab:mortality_error} displays how the related
$\rm{CV}_{Z}^{2}$ can be decomposed into separate terms, each arising
from the variances associated to the components in the definition of
$Z$. For concreteness, from Fig.~\ref{fig:boxplot}(e) we obtain
$\rm{CV}^2_{\rm IFR} ={\Sigma_{\rm IFR}^2/ \widebar{\mathrm{IFR}}}^2
\approx 1.16$ which allows us to place an upper bound on
$\sigma_{b}^{2}$ using Eq.~\eqref{sigma_f}, the results of
Tab.~\ref{tab:mortality_error}, and
\begin{equation}
\sigma_{b}^{2} < \frac{(\tilde{f}_{\rm b}-{\rm
    FPR})^{2}}{\hat{f}^{2}(1-\hat{f})^{2}}{\rm CV}^{2}_{\rm IFR} 
\approx \frac{(\tilde{f}_{\rm b}-{\rm
    FPR})^{2}}{\hat{f}^{2}(1-\hat{f})^{2}} 1.16 
\end{equation}
or on $\sigma_{\rm I}^{2}$ using $(1-\hat{f})^{2}\sigma_{\rm I}^{2} <
(\tilde{f}_{\rm b}-{\rm FPR})^{2}{\rm CV}^{2}_{\rm IFR}$.

\begin{figure}
\includegraphics{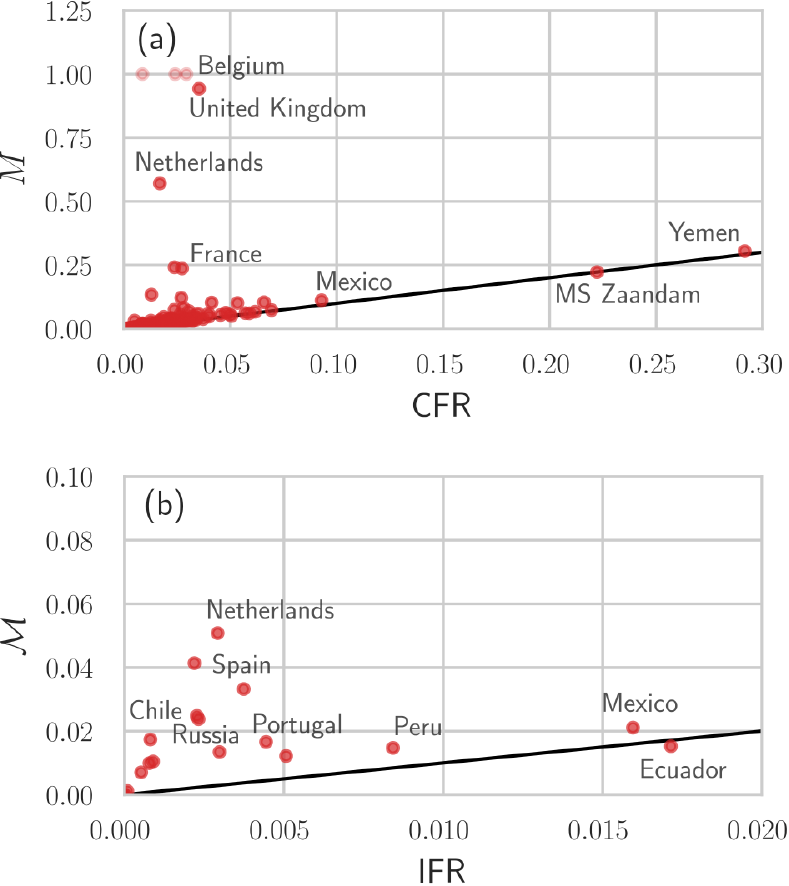}
\caption{\textbf{Different mortality measures across different
    regions.} We show the values of $M$ and $\mathrm{CFR}$ (a) and
  $\mathcal{M}$ (using $\gamma=100$) and $\mathrm{IFR}= p
  \bar{D}_{\mathrm{e}}/N_{\rm c}$ with $p=N_{\rm c}/(N_{\rm c}+N_{\rm
    u}) = 0.1$~\cite{chow2020global} (b) for different regions. The
  black solid lines have slope 1. If jurisdictions do not report the
  number of recovered individuals, $R_{\rm c} = 0$ and $M =1$ [light
    red disks in (a)].  In jurisdictions for which the data indicate
  $\bar{D}_{\rm e} < D_{\rm c}$, we set $\gamma(\bar{D}_{\rm e} -
  D_{\rm c}) = 0$ in the denominator of ${\cal M}$ which prevents it
  from becoming negative as long as $\bar{D}_{\rm e} \geq 0$. All data
  were updated on December 10,
  2020~\cite{ExcessDeathsEconomist,ExcessDeathsCDC,NYCData,dong2020interactive}.}
\label{fig:mortality2} 
\end{figure}

Finally, to provide more insight into the correlations between
different mortality measures, we plot $M$ against $\mathrm{CFR}$ and
$\mathcal{M}$ against $\mathrm{IFR}$ in Fig.~\ref{fig:mortality2}.
For most regions, we observe similar values of $M$ and $\mathrm{CFR}$
in Fig.~\ref{fig:mortality2}(a). Althouigh we expect $M \to
\mathrm{CFR}$ and $\mathcal{M} \to \mathrm{IFR}$ towards the end of an
epidemic, in some regions such as the UK, Sweden, Netherlands, and
Serbia, $M \gg {\rm CFR}$ due to unreported or incomplete reporting of
recovered cases. About 50\% of the regions that we show in
Fig.~\ref{fig:mortality2}(b) have an $\mathrm{IFR}$ that is
approximately equal to $\mathcal{M}$. Again, for regions such as
Sweden and the Netherlands, $\mathcal{M}$ is substantially larger than
$\mathrm{IFR}$ because of incomplete reporting of recovered cases.
\section*{Discussion}
\label{sec:conclusion}
\subsection*{Relevance}
In the first few weeks of the initial COVID-19 outbreak in March and
April 2020 in the US, the reported death numbers captured only about
two thirds of the total excess deaths~\cite{woolf2020excess}. This
mismatch may have arisen from reporting delays, attribution of
COVID-19 related deaths to other respiratory illnesses, and secondary
pandemic mortality resulting from delays in necessary treatment and
reduced access to health care~\cite{woolf2020excess}. We also observe
that the number of excess deaths in the Fall months of 2020 have been
significantly higher than the corresponding reported COVID-19 deaths
in many US states and countries. The weekly numbers of deaths in
regions with a high COVID-19 prevalence were up to 8 times higher than
in previous years. Among the countries that were analyzed in this
study, the five countries with the largest numbers of excess deaths
since the beginning of the COVID-19 outbreak (all numbers per 100,000)
are Peru (256), Ecuador (199), Mexico (151), Spain (136), and Belgium
(120). The five countries with the lowest numbers of excess deaths
since the beginning of the COVID-19 outbreak are Denmark (2), Norway
(6), Germany (8), Austria (31), and Switzerland
(33)~\cite{ExcessDeathsEconomist} \footnote{Note that Switzerland
  experienced a rapid growth in excess deaths in recent weeks. More
  recent estimates of the number of excess deaths per 100,000 suggest
  a value of 64~\cite{excessdata}, which is similar to the
  corresponding excess death value observed in Sweden.}. If one
includes the months before the outbreak, the numbers of excess deaths
per 100,000 in 2020 in Germany, Denmark, and Norway are -3209, -707,
and -34, respectively. In the early stages of the COVID-19 pandemic,
testing capabilities were often insufficient to resolve
rapidly-increasing case and death numbers. This is still the case in
some parts of the world, in particular in many developing
countries~\cite{ondoa2020covid}. Standard mortality measures such as
the IFR and CFR thus suffer from a time-lag problem.

%
\subsection*{Strengths and limitations}
The proposed use of excess deaths in standard mortality measures may
provide more accurate estimates of infection-caused deaths, while
errors in the estimates of the fraction of infected individuals in a
population from testing can be corrected by estimating the testing
bias and testing specificity and sensitivity.
%
%
One could sharpen estimates of the true
COVID-19 deaths by systematically analyzing the statistics of deaths
from all reported causes using a standard protocol such as
ICD-10~\cite{CDCICD10}.  
%
%
For example, the mean traffic deaths per month in Spain between
2011-2016 is about 174 persons~\cite{EUTransport}, so any
pandemic-related changes to traffic volumes would have little impact
considering the much larger number of COVID-19 deaths.

Different mortality measures are sensitive to different sources of
uncertainty. Under the assumption that all excess deaths are caused by
a given infectious disease (\eg, COVID-19), the underlying error in
the determined number of excess deaths can be estimated using
historical death statistics from the same jurisdiction. Uncertainties
in mortality measures can also be decomposed into the uncertainties of
their component quantities, including the positive-tested fraction $f$
that depend on uncertainties in the testing parameters.

As for all epidemic forecasting and surveillance, our methodology
depends on the quality of excess death and COVID-19 case data and
knowledge of testing parameters. For many countries, the lack of
binding international reporting guidelines, testing limitations, and
possible data tampering~\cite{aljazeera_deaths} complicates the
application of our framework. A striking example of variability is the
large discrepancy between excess deaths $D_{\rm e}$ and confirmed
deaths $D_{\rm c}$ across many jurisdictions which render mortalities
that rely on $D_{\rm c}$ suspect. More research is necessary to
disentangle the excess deaths that are directly caused by SARS-CoV-2
infections from those that result from postponed medical
treatment~\cite{woolf2020excess}, increased suicide
rates~\cite{sher2020impact}, and other indirect factors contributing
to an increase in excess mortality. Even if the numbers of excess
deaths were accurately reported and known to be caused by a given
disease, inferring the corresponding number of unreported cases (\eg,
asymptomatic infections), which appears in the definition of the IFR
and $\mathcal{M}$ (see Tab.~\ref{tab:mortality_measures}), is
challenging and only possible if additional models and assumptions are
introduced.

Another complication may arise if the number of excess deaths is not
significantly larger than the historical mean. Then,
excess-death-based mortality estimates suffer from large
uncertainty/variability and may be meaningless.  While we have
considered only the average or last values of $\tilde{f}_{\rm b}$, our
framework can be straightforwardly extended and dynamically applied
across successive time windows, using \textit{e.g.}, Bayesian or
Kalman filtering approaches.

Finally, we have not resolved the excess deaths or mortalities with
respect to age or other attributes such as sex, co-morbidities,
occupation, etc. We expect that age-structured excess deaths better
resolve a jurisdiction's overall mortality.  By expanding our testing
and modeling approaches on stratified data, one can also
straightforwardly infer stratified mortality measures $Z$, providing
additional informative indices for comparison.
\section*{Conclusions} 
Based on the data presented in Figs.~\ref{fig:boxplot} and
\ref{fig:mortality2}, we conclude that the mortality measures $r$,
$\mathrm{CFR}$, $\mathrm{IFR}$, $M$, and $\mathcal{M}$ may provide
different characterizations of disease severity in certain
jurisdictions due to testing limitations and bias, differences in
reporting guidelines, reporting delays, etc. The propagation of
uncertainty and coefficients of variation that we summarize in
Tab.~\ref{tab:mortality_error} can help quantify and compare errors
arising in different mortality measures, thus informing our
understanding of the actual death toll of COVID-19.  Depending on the
stage of an outbreak and the currently available disease monitoring
data, certain mortality measures are preferable to others. If the
number of recovered individuals is being monitored, the resolved
mortalities $M$ and $\mathcal{M}$ should be preferred over
$\mathrm{CFR}$ and $\mathrm{IFR}$, since the latter suffer from errors
associated with the time-lag between infection and
resolution~\cite{bottcher2020case}. For estimating $\mathrm{IFR}$ and
$\mathcal{M}$, we propose using excess death data and an epidemic
model. In situations in which case numbers cannot be estimated
accurately, the relative excess deaths $r$ provides a complementary
measure to monitor disease severity. Our analyses of different
mortality measures reveal that

\begin{itemize}

\item The CFR and $M$ are defined directly from confirmed deaths $D_{\rm
  c}$ and suffers from variability in its reporting. Moreover,
  the CFR does not consider resolved cases and is expected to evolve
  during an epidemic. Although $M$ includes resolved cases, its
  additionally required confirmed recovered cases $R_{\rm c}$ add
  to its variability across jurisdictions.  Testing errors affect both
  $D_{\rm c}$ and $R_{\rm c}$, but if the FNR and FPR are known, they
  can be controlled using Eq.~\eqref{PTOT_APP} given in the SI.

\item The IFR requires knowledge of the true cumulative number of
  disease-caused deaths as well as the true number of infected
  individuals (recovered or not) in a population. We show how these
  can be estimated from excess deaths and testing, respectively. Thus,
  the IFR will be sensitive to the inferred excess deaths and from the
  testing (particularly from the bias in the testing). Across all
  countries analyzed in this study, we found a mean IFR of about
  0.24\% (95\% CI 0.0--1.5\%), which is similar to the previously
  reported values between 0.1 and
  1.5\%~\cite{salje2020estimating,chow2020global,ioannidis2020infection}.

\item In order to estimate the resolved true mortality ${\cal M}$, an
  additional relationship is required to estimate the unconfirmed
  recovered population $R_{\rm u}$. In this paper, we propose a simple
  SIR-type model in order to relate $R_{\rm u}$ to measured excess and
  confirmed deaths through the ratio of the recovery rate to the death
  rate. The variability in reporting $D_{\rm c}$ across different
  jurisdictions generates uncertainty in ${\cal M}$ and reduces its
  reliability when compared across jurisdictions.

\item The mortality measures that can most reliably be compared across
  jurisdictions should not depend on reported data which are subject
  to different protocols, errors, and manipulation/intentional
  omission. Thus, the per capital excess deaths and relative excess
  deaths $r$ (see last column of Table \ref{tab:mortality_measures})
  are the measures that provide the most consistent comparisons of
  disease mortality across jurisdictions (provided total deaths are
  accurately tabulated). However, they are the least informative in
  terms of disease severity and individual risk, for which $M$ and
  ${\cal M}$ are better.

\item Uncertainty in all mortalities $Z$ can be decomposed into the
  uncertainties in component quantities such as the excess death
  or testing bias. We can use global data to estimate the means and
  variances in $Z$, allowing us to put bounds on the variances of the
  component quantities and/or parameters.  
%

\end{itemize}

Parts of our framework can be readily integrated into or combined with
mortality surveillance platforms such as the European Mortality
Monitor (EURO MOMO) project~\cite{euromomo} and the Mortality
Surveillance System of the National Center for Health
Statistics~\cite{USData} to assess disease burden in terms of
different mortality measures and their associated uncertainty.
\section*{Data availability}
All datasets used in this study are available from
Refs.~\cite{USData,SpainData,EnglandData,SwitzerlandData,Istat}. The
source codes used in our analyses are publicly available at
\cite{GitHub}.
\section*{Acknowledgements}
LB acknowledges financial support from the Swiss National Fund
(P2EZP2\_191888). The authors also acknowledge financial support from
the Army Research Office (W911NF-18-1-0345), the NIH (R01HL146552),
and the National Science Foundation (DMS-1814364, DMS-1814090).
\onecolumngrid
\bibliography{refs}
\bibliographystyle{unsrtnat}
\renewcommand{\theequation}{A\arabic{equation}}
\renewcommand{\thefigure}{A\arabic{figure}} 
\setcounter{figure}{0}   
\setcounter{equation}{0}  

\onecolumngrid

\newpage
\section*{Supplementary Information}

\subsection*{Examples of excess death data}

\begin{figure}[h!]
\centering
\includegraphics[width = 0.75\textwidth]{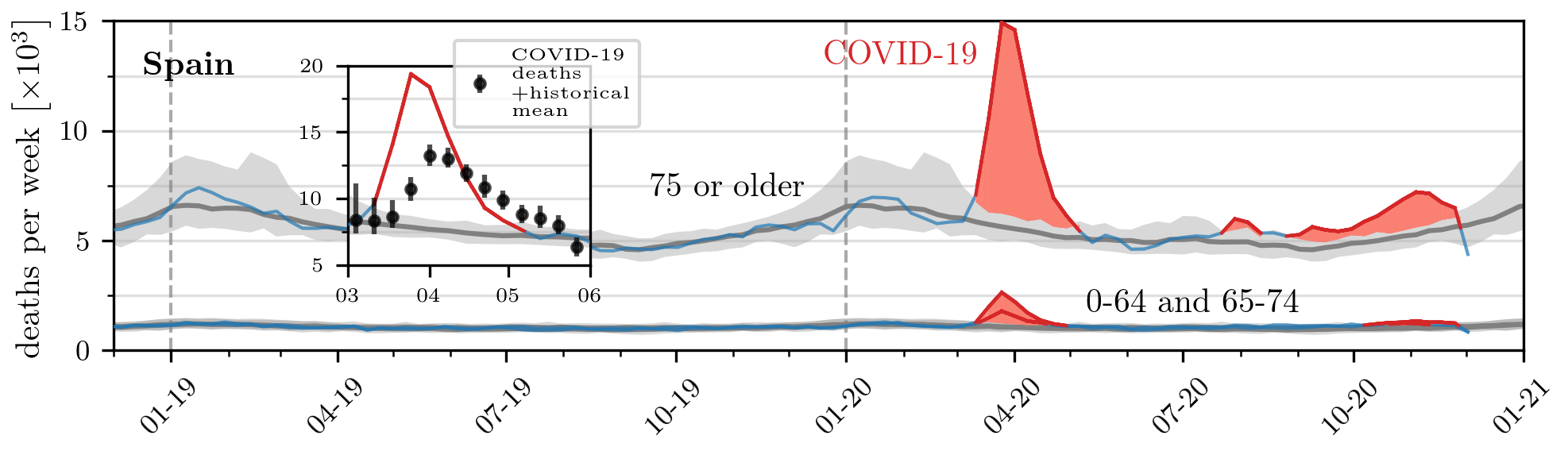}
\includegraphics[width = 0.75\textwidth]{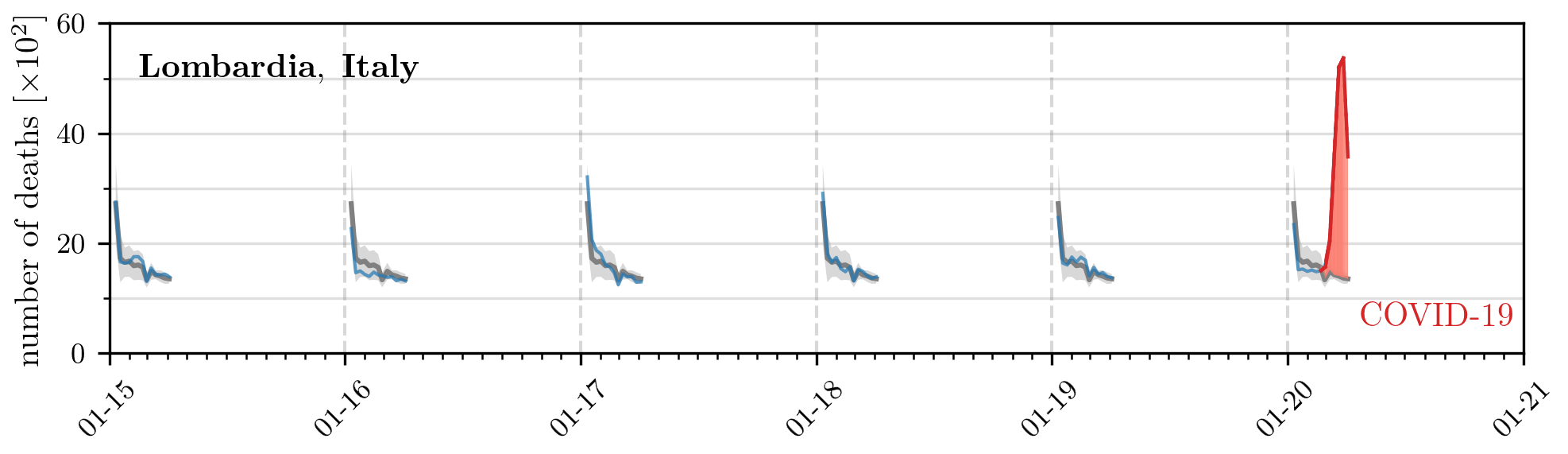}
\includegraphics[width = 0.75\textwidth]{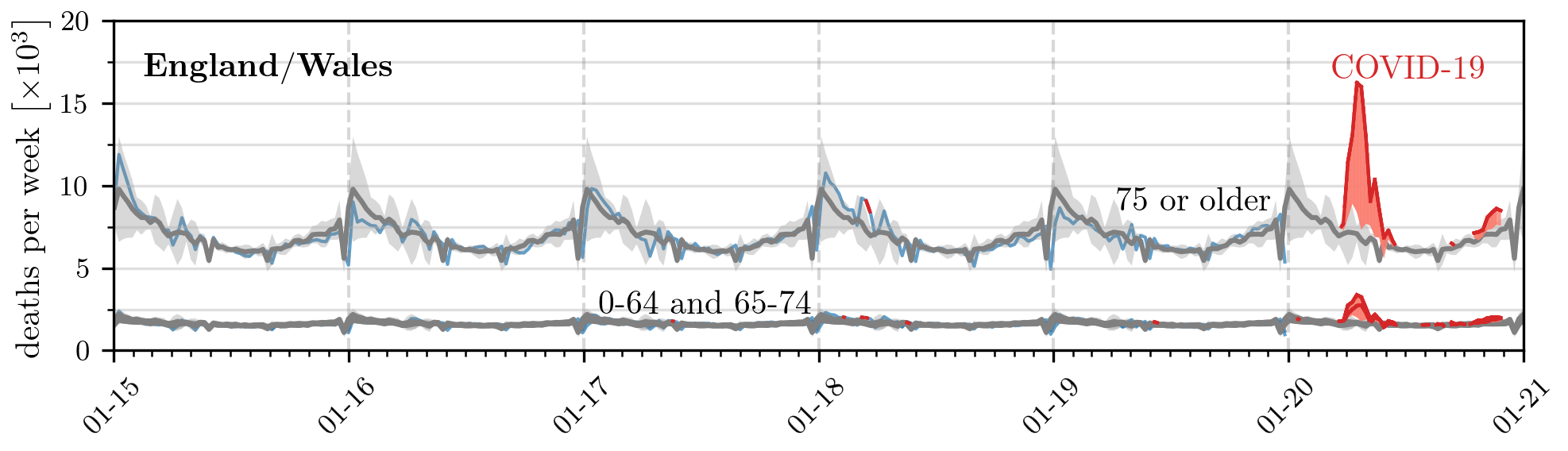}
\includegraphics[width = 0.75\textwidth]{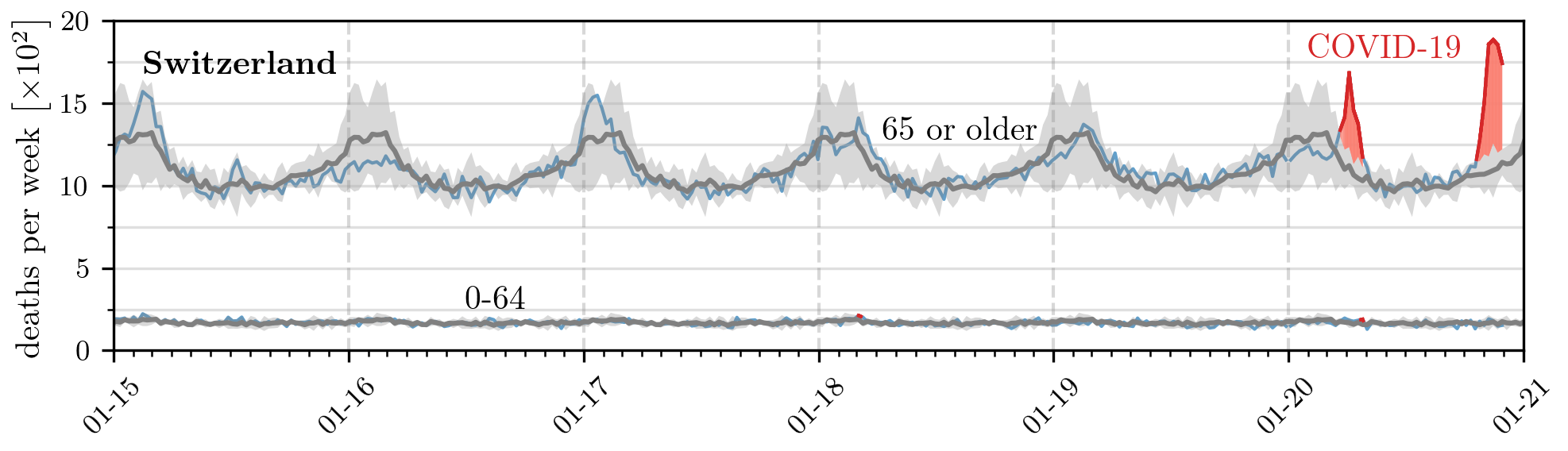}
\caption{\textbf{Mortality evolution in different countries}. The
  evolution of weekly deaths in New York City, Spain, England/Wales,
  and Switzerland for different age classes (where available). Grey
  solid lines and shaded regions represent the historical mean numbers
  of deaths and corresponding confidence intervals. Blue solid lines
  indicate weekly deaths and weekly deaths that lie outside the
  confidence intervals are indicated by solid red lines. For
  England/Wales and Switzerland, weekly means and 95\% confidence
  intervals are based on data from 2015--2019. In the case of Spain,
  we show the reported COVID-19 deaths across all age
  classes~\cite{AllReps} in the inset and use the 99\% confidence
  intervals that are directly provided in the corresponding
  data~\cite{excessdata}. The red shaded regions represent the mean
  cumulative excess deaths $D_{\rm e}$. The data are derived from
  Refs.~\cite{USData,SpainData,EnglandData,SwitzerlandData,Istat}.}
\label{fig:panels}
\end{figure}

We tally weekly deaths according to Eq.\,\eqref{DK_STATS}
for each week $i$ starting from the first week of 2020, and cumulative
excess deaths as in Eq.\,\eqref{DE_STATS} adding all weekly
contributions from the first week of 2020 onwards.  Note that some
governmental agencies tabulate weekly deaths starting on the Sunday
closest to January 1 2020 (December 29 2019, such as the United
States), others instead use January 1 2020 as the first day of the
week (such as Germany).  A detailed list of how each country bins
weekly deaths is included in Ref.\,\cite{ExcessDeathsEconomist}.  The
final week $k$ up to which the cumulative count is taken depends on
data availability, since some countries have larger reporting delays
than others. In the majority of cases $k$ is beyond the fourth week of
November 2020. Quantities are calculated from data that include
deaths from typically $J=5$ previous years~\cite{ExcessDeathsEconomist}.
\begin{figure}[tp!]
\centering
\includegraphics[width = 0.45\textwidth]{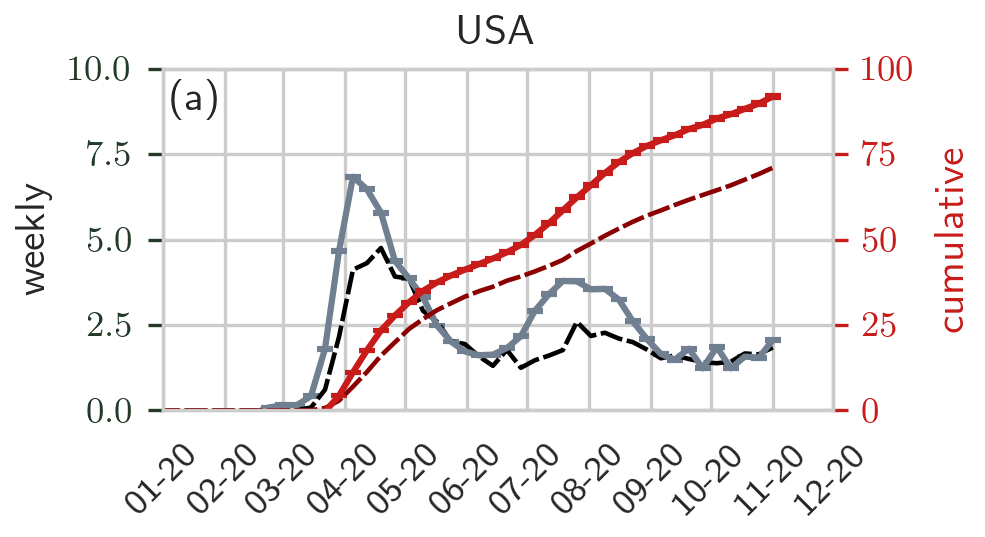}
\hspace{0.8cm}
\includegraphics[width = 0.45\textwidth]{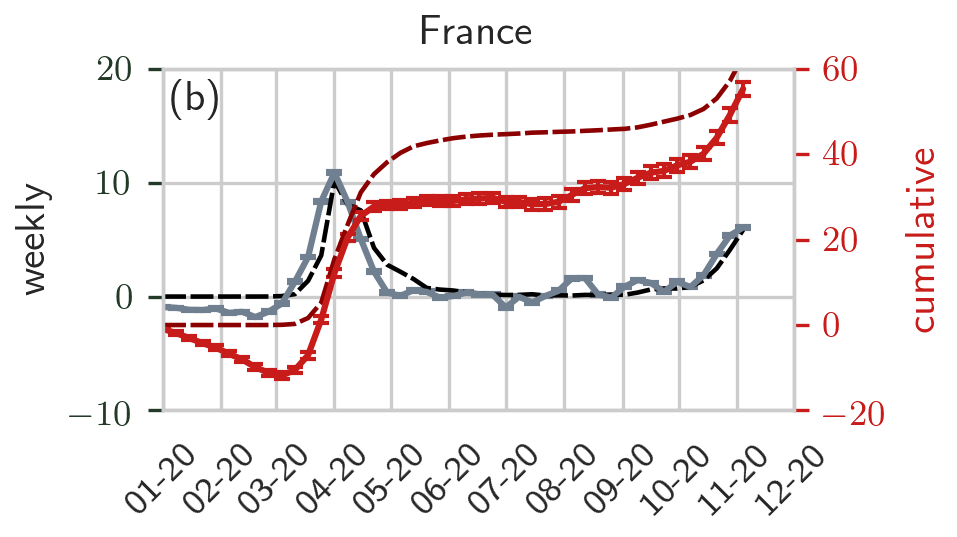}
\includegraphics[width = 0.45\textwidth]{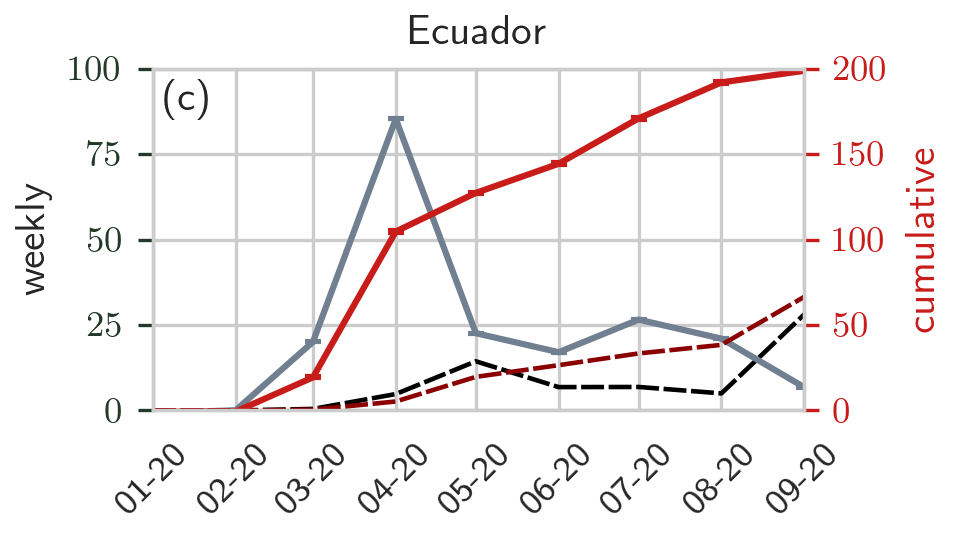}
\hspace{0.8cm}
\includegraphics[width = 0.45\textwidth]{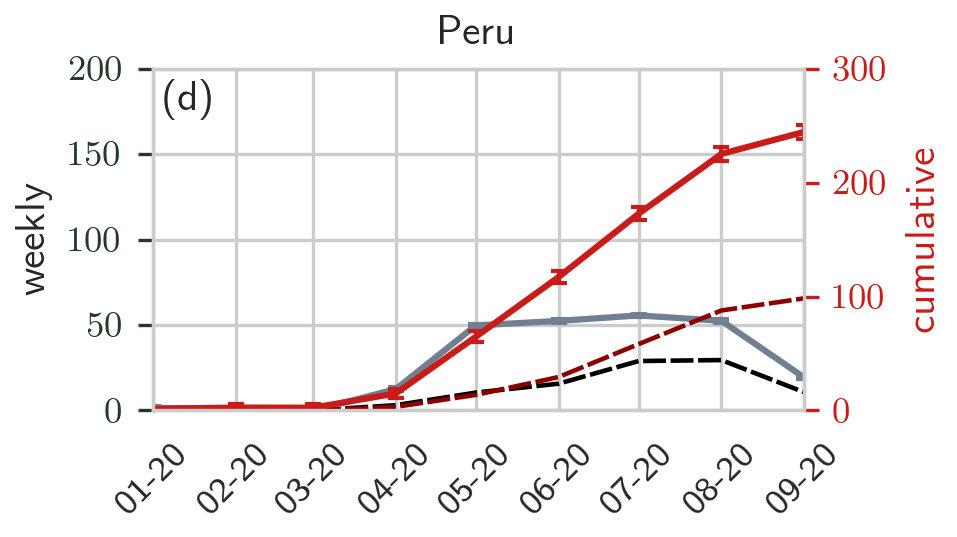}
\includegraphics[width = 0.45\textwidth]{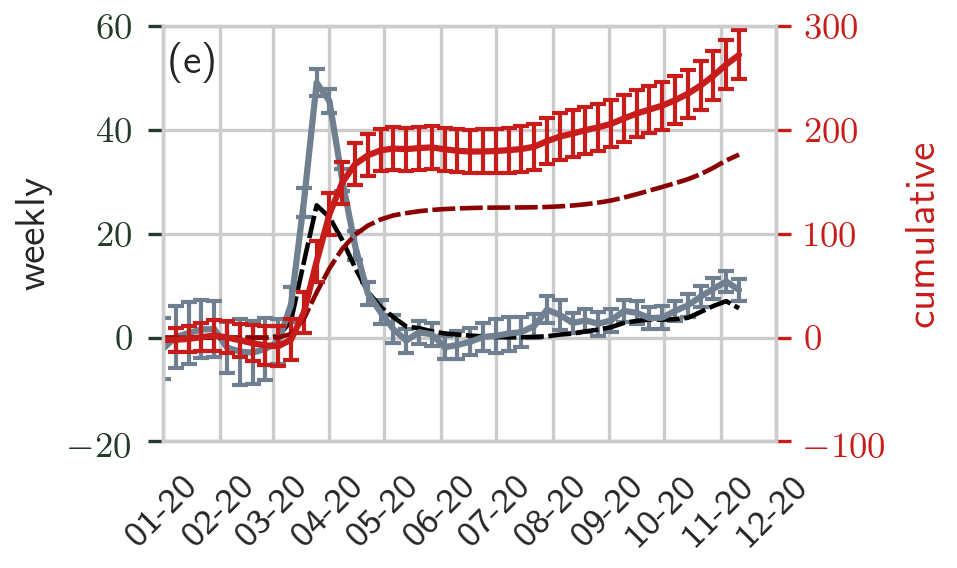}
\hspace{0.8cm}
\includegraphics[width = 0.45\textwidth]{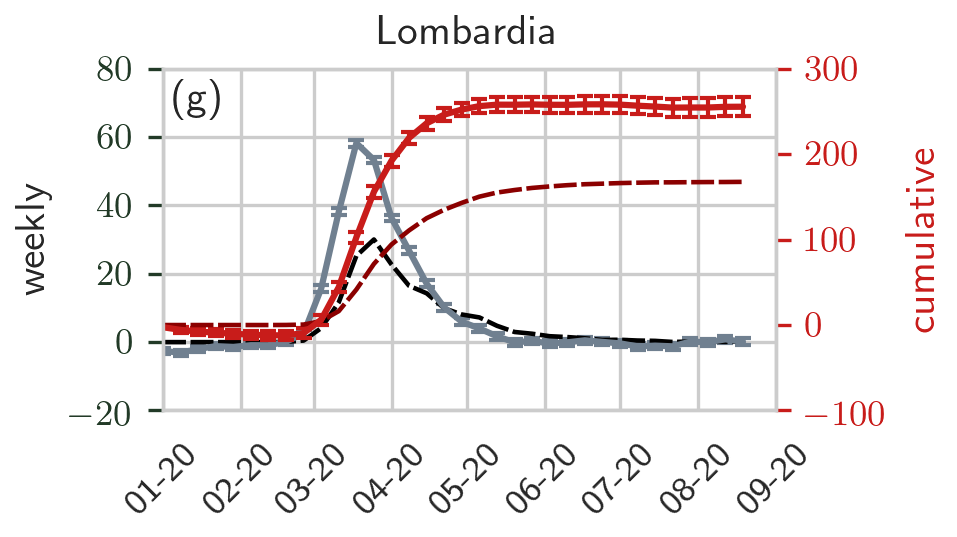}
\includegraphics[width = 0.45\textwidth]{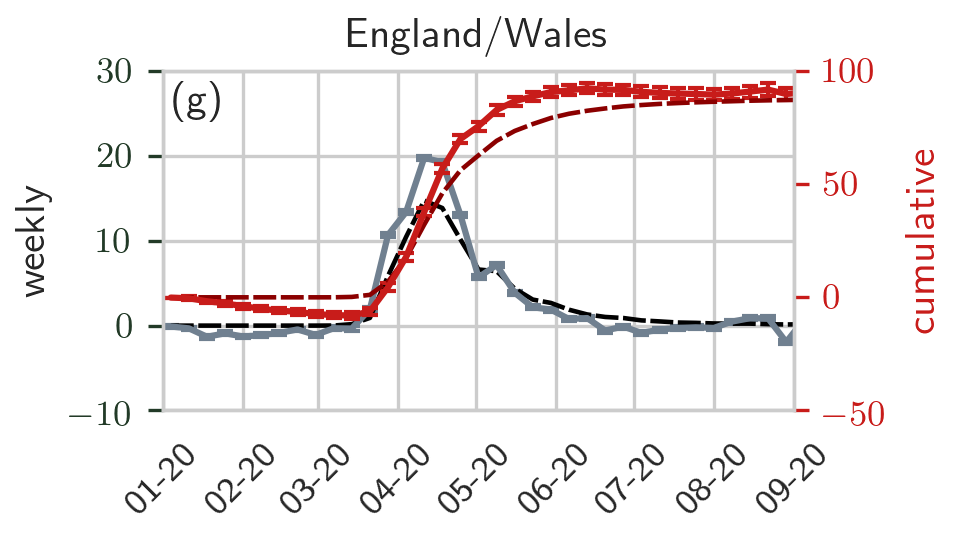}
\hspace{0.8cm}
\includegraphics[width = 0.45\textwidth]{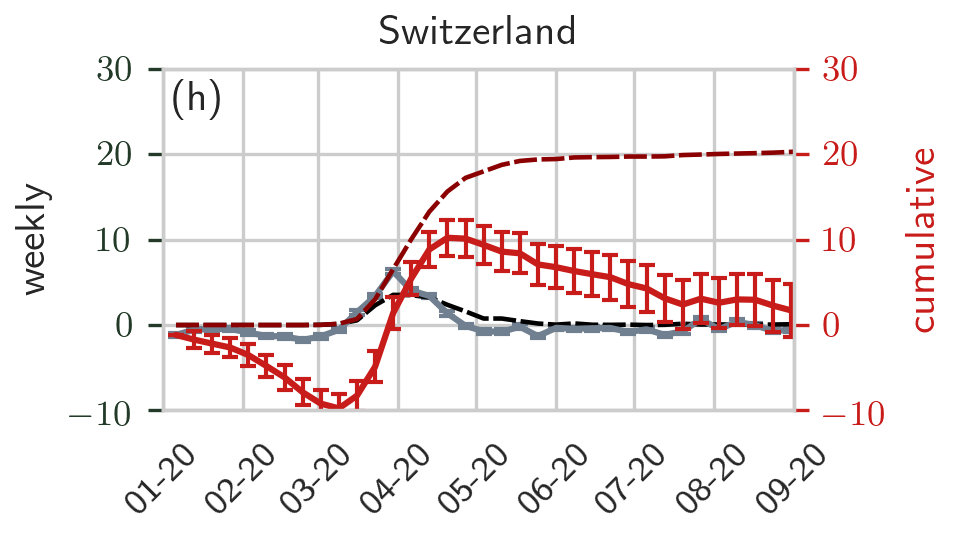}

\caption{\textbf{Weekly and cumulative death rates in different
    countries and regions}. We compare the evolution of confirmed
  weekly deaths $d^{(0)}_{\rm c}(i)$ (dashed black curves) and
  cumulative deaths $D_{\rm c}(k)$ (dashed dark red curves) with
  weekly excess deaths $\bar d_{\rm e}(i)$ (solid grey curves) and
  cumulative excess deaths $\bar{D}_{\rm e}(k)$ (solid red
  curves). The deaths are plotted in units of per 100,000 in different
  countries and regions.  The data are derived from
  Ref.~\cite{ExcessDeathsEconomist} and the error bars for the excess
  deaths are derived from Eqs.~\eqref{DK_STATS} and
  \eqref{DE_STATS}. For Spain, we used the 99\% confidence intervals
  that are directly provided in the corresponding
  data~\cite{excessdata} to approximate the 95\% confidence
  intervals. Typically, we find $\bar{D}_{\rm e}(k) >D_{\rm c}(k)$.}
\label{fig:excess_rate}
\end{figure}
In Fig.~\ref{fig:excess_rate} we plot the weekly confirmed deaths
$d_{\rm c}^{(0)}(i)$, the cumulative deaths $D_{\rm c}(k) =
\sum_{i=1}^k d^{(0)}_{\rm c}(i)$, and the mean weekly and cumulative
excess deaths $\bar{d}_{\rm e}(i)$ for 2020 as available from data.
We also show $\bar{D}_{\rm e}(k)$ per 100,000 persons from the start
of 2020 using Eqs.~\eqref{DK_STATS} and \eqref{DE_STATS}.  The
corresponding error bars in Fig.\,\ref{fig:excess_rate} indicate 95\%
confidence intervals defined by $\bar{d}_{\rm e}(i)\pm 1.96
\,\sigma_{\rm e}(i)$ and $\bar{D}_{\rm e}(k)\pm 1.96 \, \Sigma_{\rm
  e}(k)$ in Eqs.~\eqref{DK_STATS} and \eqref{DE_STATS}, respectively.
For Spain, we used the 99\% confidence intervals that are directly
provided in the corresponding data~\cite{excessdata} to approximate
the 95\% confidence intervals. Excess death statistics evolve
differently across different countries and regions. For example, in
France excess deaths were negative until the end of March 2020,
quickly increasing in April 2020. In Ecuador and Peru, the number of
excess deaths is more than 2.5 times larger than the corresponding
number of confirmed COVID-19 deaths.

\subsection*{Statistical testing model}

Given biases in sampling and testing errors, it is important to use a
statistical testing model that takes them into account when estimating
the fraction $f$ of a population $N$ that are infected.  Testing
biases arises, for example, if symptomatic individuals are more likely
to seek testing. Thus, the probability $f_{\rm b}$ that an individual
who chooses to be tested is positive may be different from $f$ the
probability that a \textit{randomly} selected individual is positive,
as defined in Eq.~\eqref{F_B}.  If all tests are error-free, the
probability that $Q^{+}$ positive results arise from the $Q \geq
Q^{+}$ administered tests is given by
\begin{equation}
P_{\rm true}(Q^{+}\vert Q, f_{\rm b}) = 
{Q \choose Q^{+}} f_{\rm b}^{Q^{+}}(1-f_{\rm b})^{Q-Q^{+}}.
\label{P_EXACT_BINOMIAL}
\end{equation}

Eq.~\eqref{P_EXACT_BINOMIAL} is derived under the assumption that once
individuals are tested, they are ``replaced'' in the population and
can be tested again. The analogous distribution $P_{\rm
  true}(Q^{+}\vert Q, f_{\rm b})$ for testing ``without replacement''
can be straightforwardly derived and yields results quantitatively
close to Eq.~\eqref{P_EXACT_BINOMIAL} provided $Q/N \lesssim 0.3$.

Eq.~\eqref{P_EXACT_BINOMIAL} also assumes flawless testing.  Tests with
Type I (false positives) and Type II (false negatives) may wrongly
catalog uninfected individuals as infected (with rate FPR) while
missing some infected individuals (with rate FNR). For serological
COVID-19 tests, such as antibody tests, the estimated percentages of
false positives and false negatives are typically low, with
$\mathrm{FPR}\approx 0.03-0.07$ and $\mathrm{FNR}\approx
0.1$~\cite{fda_serological,cohen2020diagnosing,watson2020interpreting}.
For RT-PCR tests, the FNRs depend strongly on the actual assay
method~\cite{lassauniere2020evaluation,whitman2020test} and typically
lie between 0.1 and 0.3~\cite{fang2020sensitivity,wang2020detection}
but might be as high as $\mathrm{FNR}\approx 0.68$ if throat swabs are
used~\cite{wang2020detection,watson2020interpreting}. FNRs can also
vary significantly depending on how long after initial infection the
test is administered~\cite{bastos2020diagnostic}.  A systematic review
conducted worldwide found ${\rm FNR}\approx 0.54$ at initial
testing~\cite{arevalo2020false}, underlying the need for retesting.
Reported percentages of false positives in RT-PCR tests are about
$\mathrm{FPR}\approx 0.05$~\cite{watson2020interpreting}. A large
meta-analysis of serological tests estimates $\mathrm{FPR}\approx
0.02$ and $\mathrm{FNR}\approx 0.02-0.16$~\cite{bastos2020diagnostic}.
These testing errors can lead to inaccurate estimates of disease
prevalence; uncertainty in FPR, FNR will thus lead to uncertainty in
the estimate of prevalence.

As illustrated through Fig.~\ref{fig:metrics}, errors in testing may
result in the recorded number $\tilde{Q}^{+}$ of positive tests to be
different from the $Q^{+}$ that would be obtained under perfect
testing. The probability that $\tilde{Q}^{+}$ positive tests are
returned due to testing errors can be described in terms of $Q^{+}$,
${\rm FPR}$, and ${\rm FNR}$ and the corresponding probability
distribution $P_{\rm err}(\tilde{Q}^{+}\vert Q^{+}, \rm{FPR},
\rm{FNR})$ is given by

\begin{equation}
P_{\rm err}(\tilde{Q}^{+}\vert Q^{+},{\rm FPR},{\rm FNR}) = 
\sum_{p_{+}=0}^{\tilde{Q}^{+}}
{Q^{+} \choose p_{+}} (1-{\rm FNR})^{p_{+}} ({\rm FNR})^{Q^{+}-p_{+}}
{Q^{-} \choose q_{+}} ({\rm FPR})^{q_{+}}(1-{\rm FPR})^{Q^{-}-q_{+}}.
\label{P_ERR}
\end{equation}
where $q_{+}\equiv \tilde{Q}^{+}-p_{+}$. By convolving $P_{\rm
  err}(\tilde{Q}^{+} \vert Q^{+}, \rm{FPR}, \rm{FNR})$ with $P_{\rm
  true} (Q^+ \vert Q, f_{\rm b})$ we derive the overall likelihood
distribution for the measured number $\tilde{Q}^{+}$ of true
\textit{and} false positives given a set of specified parameters
$\theta=\{Q, f, b, {\rm FPR},{\rm FNR}\}$ describing the population
and testing

\begin{equation}
P(\tilde{Q}^{+}\vert Q, f, b, {\rm FPR},{\rm
  FNR}) = \sum_{Q^{+}=0}^{Q} P_{\rm err}(\tilde{Q}^{+}\vert Q^{+},
{\rm FPR}, {\rm FNR}) P_{\rm true}(Q^{+}\vert Q, f_{\rm b}(f, b)).
\label{PTOT_APP}
\end{equation}
When $Q^{+}, \tilde{Q}^{+}$, and $Q \gg1$, we can approximate
$P_{\rm true}$, $P_{\rm err}$, and $P$ by normal
distributions and rewrite $P$ as a function of the observed
positive fraction $\tilde{f}_{\rm b} \equiv \tilde{Q}^{+}/Q$
(Eqs.~\eqref{PTOT} and \eqref{MU_SIGMA}).


Using Bayes' rule, we can then formally define the likelihood of
$\theta$ given a measured $\tilde{f}_{\rm b}$,
\begin{equation}
P(\theta \vert \tilde{f}_{\rm b}, \alpha) = 
{P(\tilde{f}_{\rm b}\vert\theta)P_{0}(\theta\vert \alpha)\over
  \sum_{\theta} P(\tilde{f}_{\rm b}\vert \theta)P_{0}(\theta\vert
  \alpha)},
\label{BAYES}
\end{equation}
where $\alpha =\{\bar{\theta}, \sigma_{\theta}\}$ are hyperparameters
defining the prior $P_{0}(\theta\vert \alpha)$, such as their means
$\bar{\theta}=\{\bar{D}_{\rm e}, \widebar{{\rm FPR}}, \widebar{{\rm
    FNR}}, \bar{b}, \bar{N}\}$ and standard deviations
$\sigma_{\theta}=\{\Sigma_{\rm e}, \sigma_{\rm I}, \sigma_{\rm II}, b,
\sigma_{b}, \sigma_{N}\}$. Formally, the probability of measuring a
value of a mortality measure $Z={\rm CFR}, {\rm IFR}, M, {\cal M}$, or
$r$, can be computed from

\begin{equation}
P(Z\vert \alpha) = \int P(Z\vert \theta)P(\theta\vert \alpha)\dd \theta,
\end{equation}
where $P(Z\vert \theta)$ defines the statistical model of the
mortality measure given the components and parameters $\theta$ and the
hyperparameters $\alpha$ defining the distribution over $\theta$. For
example, if $Z$ is the value of the IFR, $\theta = \{D_{\rm e}, f,
N\}$ and $\alpha = \{(\bar{D}_{\rm e}, \Sigma_{\rm e}), (\bar{b},
\widebar{{\rm FPR}}, \widebar{{\rm FNR}}, \sigma_{b}, \sigma_{\rm I},
\sigma_{\rm II}), (\bar{N}, \Sigma_{N})\}$ are the mean and standard
deviation of excess deaths, testing parameters, and the total
population, respectively.

A simpler way to incorporate uncertainty in the infected fraction $f$
is to assume a Gaussian approximation for all distributions and
propagate the uncertainty in testing parameters. The squared
coefficient of variation ${\rm CV}_{f}^{2}$ is then decomposed into
the parameter variances according to

\begin{equation}
{\sigma_{f}^{2}\over \hat{f}^{2}} 
\approx {(1-(1-e^{b})\hat{f})^{2}\over X^{2} Q}
\tilde{f}_{\rm b}(1-\tilde{f}_{\rm b})+{(1-\hat{f})^{2}\over X^{2}}\sigma_{\rm I}^{2} + 
{e^{2b}\hat{f}^{2}\over X^{2}}\sigma_{\rm II}^{2} +{\hat{f}^{2}(1-\hat{f})^{2}\over X^{2}}
\sigma_{b}^{2},
\label{sigma_f}
\end{equation}
where $X\equiv \tilde{f}_{\rm b}-{\rm FPR}$. The values of $b$, ${\rm
  FPR}$, ${\rm FNR}$ above are mean or maximum likelihood estimates of
the bias and testing errors, and $\sigma_{b}^{2}$, $\sigma_{\rm
  I}^{2}$, and $\sigma_{\rm II}^{2}$ are their associated
uncertainties.  The means and variances $\{\bar{b}, \widebar{{\rm
    FPR}}, \widebar{{\rm FNR}}, \sigma_{b}^{2}, \sigma_{\rm I}^{2},
\sigma_{\rm II}^{2}\}$ represent hyperparameters associated with
testing (see SI). Our result for $\sigma_{f}^{2}$ in
Eq.~\eqref{sigma_f} assumes $\{b, {\rm FPR}, {\rm FNR}\}$ are
uncorrelated.  Since $Q\gg 1$ is typically large, we expect the first
contribution to the uncertainty, arising from stochasticity in the
sampling and proportional to $\tilde{f}_{\rm b}(1-\tilde{f}_{\rm
  b})/Q$ to be negligible. Uncertainties in other quantities will
ultimately contribute to uncertainty in the mortalities $Z$, as listed
in Table~\ref{tab:mortality_error}.

\subsection*{Modeling of resolved mortality}
\label{app:mortality}
\begin{figure}
\includegraphics{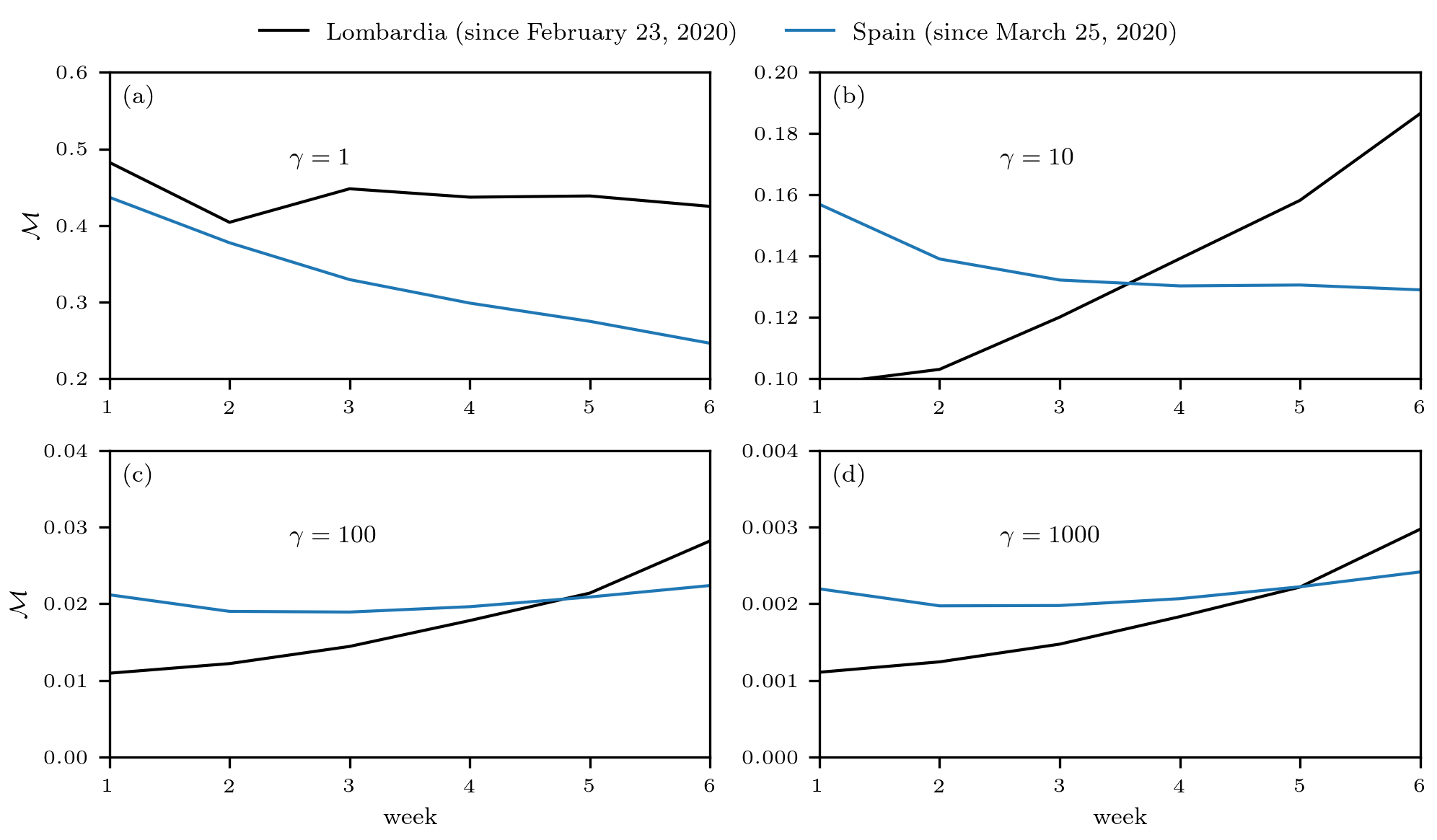}
\caption{\textbf{Evolution of resolved mortality.} We show the
  evolution of $\mathcal{M}(t)$ for different values of effective
  recovery rates of unreported cases $\gamma$. The data are derived
  from Refs.~\cite{Istat,SpainData}.}
\label{fig:resolved_mortality}
\end{figure}

In Fig.~\ref{fig:resolved_mortality}, we show the evolution of
$\mathcal{M}$ for Spain and Lombardia, using different effective
recovery rates of unreported cases $\gamma$. We compute $\mathcal{M}$
according to Eq.~\eqref{eq:Mp_2} and use excess mortality data of
Fig.~\ref{fig:panels} to determine $\bar{D}_{\rm e}$. The
corresponding data for confirmed recovered and deceased individuals,
$R_{\rm c}$ and $D_{\rm c}$, is taken from
Ref.~\cite{excessdata}. Current estimates of the IFR are
$0.1-1.5$\%~\cite{salje2020estimating,chow2020global,ioannidis2020infection}. To
obtain a value of $\mathcal{M}$ in a similar range, we vary $\gamma$
from $1-1000$ and find that $\mathcal{M} \approx 0.1-1$\% is
consistent with $\gamma=100-1000$.
\end{document}